\documentclass[a4paper,journal,12pt,draftclsnofoot,onecolumn]{IEEEtran}
\usepackage{cite}
\ifCLASSINFOpdf
\else
\fi
\usepackage[cmex10]{amsmath}
\usepackage{units}
\usepackage{mathrsfs}
\usepackage{amssymb}
\usepackage{bm}

\usepackage{tikz}
\usetikzlibrary{arrows}

\usepackage{algorithm}

\usepackage{multirow}

\usepackage{algorithmic}
\newtheorem{thm}{Theorem}
\newtheorem{defn}{Definition}
\newtheorem{lem}{Lemma}
\newtheorem{cor}{Corollary}
\newtheorem{rmk}{Remark}

\begin{document}

% paper title
% can use linebreaks \\ within to get better formatting as desired
\title{Coded Caching in Fog-RAN: $b$-Matching Approach}

% author names and IEEE memberships
% note positions of commas and nonbreaking spaces ( ~ ) LaTeX will not break
% a structure at a ~ so this keeps an author's name from being broken across
% two lines.
% use \thanks{} to gain access to the first footnote area
% a separate \thanks must be used for each paragraph as LaTeX2e's \thanks
% was not built to handle multiple paragraphs

\author{Bo~Bai,~\IEEEmembership{Senior Member,~IEEE}, Wanyi Li,~\IEEEmembership{Student Member,~IEEE}, Li Wang,~\IEEEmembership{Senior Member,~IEEE}, Gong Zhang% <-this % stops a space

%\thanks{This work is supported in part by XXXX. A part of this work was presented at the IEEE International Conference on Communications, Budapest, Hungary, June 2013 \cite{bai_conditional_2013}.}% <-this % stops a space
\thanks{The corresponing authors of this paper are B. Bai (baibo8@huawei.com) and L. Wang (liwang@bupt.edu.cn).}% <-this % stops a space
\thanks{B. Bai and G. Zhang are with Future Network Theory Lab, 2012 Labs, Huawei Technologies Co., Ltd., Shatin, N. T., Hong Kong (e-mail: baibo8@huawei.com, nicholas.zhang@huawei.com).}% <-this % stops a space
\thanks{W. Li and L. Wang are with School of Electronic Engineering, Beijing University of Posts and Telecommunications, Beijing 100876, China (e-mail: lwy@bupt.edu.cn, liwang@bupt.edu.cn).}% <-this % stops a space
}

% note the % following the last \IEEEmembership and also \thanks -
% these prevent an unwanted space from occurring between the last author name
% and the end of the author line. i.e., if you had this:
%
% \author{....lastname \thanks{...} \thanks{...} }
%                     ^------------^------------^----Do not want these spaces!
%
% a space would be appended to the last name and could cause every name on that
% line to be shifted left slightly. This is one of those "LaTeX things". For
% instance, "\textbf{A} \textbf{B}" will typeset as "A B" not "AB". To get
% "AB" then you have to do: "\textbf{A}\textbf{B}"
% \thanks is no different in this regard, so shield the last } of each \thanks
% that ends a line with a % and do not let a space in before the next \thanks.
% Spaces after \IEEEmembership other than the last one are OK (and needed) as
% you are supposed to have spaces between the names. For what it is worth,
% this is a minor point as most people would not even notice if the said evil
% space somehow managed to creep in.

% The paper headers
\markboth{IEEE Transactions on Communications, vol. XX, no. XX, MONTH YEAR}%
{Bai \MakeLowercase{\textit{et al.}}: Coded Caching in Fog-RAN: \MakeLowercase{$b$}-Matching Approach}
% The only time the second header will appear is for the odd numbered pages
% after the title page when using the twoside option.
%
% *** Note that you probably will NOT want to include the author's ***
% *** name in the headers of peer review papers.                   ***
% You can use \ifCLASSOPTIONpeerreview for conditional compilation here if
% you desire.

% If you want to put a publisher's ID mark on the page you can do it like
% this:
%\IEEEpubid{0000--0000/00\$00.00~\copyright~2007 IEEE}
% Remember, if you use this you must call \IEEEpubidadjcol in the second
% column for its text to clear the IEEEpubid mark.

% use for special paper notices
%\IEEEspecialpapernotice{(Invited Paper)}

% make the title area
\maketitle

\begin{abstract}
	\boldmath Fog radio access network (Fog-RAN), which pushes the caching and computing capabilities to the network edge, is capable of efficiently delivering contents to users by using carefully designed caching placement and content replacement algorithms. In this paper, the transmission scheme design and coding parameter optimization will be considered for coded caching in Fog-RAN, where the reliability of content delivery, i.e., content outage probability, is used as the performance metric. The problem will be formulated as a complicated multi-objective probabilistic combinatorial optimization. A novel maximum $b$-matching approach will then be proposed to obtain the Pareto optimal solution with fairness constraint. Based on the fast message passing approach, a distributed algorithm with a low memory usage of $O(M+N)$ is also proposed, where $M$ is the number of users and $N$ is the number of Fog-APs. Although it is usually very difficult to derive the closed-form formulas for the optimal solution, the approximation formulas of the content outage probability will also be obtained as a function of coding parameters. The asymptotic optimal coding parameters can then be obtained by defining and deriving the outage exponent region (OER) and diversity-multiplexing region (DMR). Simulation results will illustrate the accuracy of the theoretical derivations, and verify the outage performance of the proposed approach. Therefore, this paper not only proposes a practical distributed Fog-AP selection algorithm for coded caching, but also provides a systematic way to evaluate and optimize the performance of Fog-RANs.
\end{abstract}
% IEEEtran.cls defaults to using nonbold math in the Abstract.
% This preserves the distinction between vectors and scalars. However,
% if the journal you are submitting to favors bold math in the abstract,
% then you can use LaTeX's standard command \boldmath at the very start
% of the abstract to achieve this. Many IEEE journals frown on math
% in the abstract anyway.

% Note that keywords are not normally used for peer review papers.
\begin{IEEEkeywords}
	Coded caching, Fog-RAN, $b$-matching, fast message passing, saddle-point method, outage exponent region.
\end{IEEEkeywords}

% For peer review papers, you can put extra information on the cover
% page as needed:
% \ifCLASSOPTIONpeerreview
% \begin{center} \bfseries EDICS Category: 3-BBND \end{center}
% \fi
%
% For peerreview papers, this IEEEtran command inserts a page break and
% creates the second title. It will be ignored for other modes.
\IEEEpeerreviewmaketitle

\section{Introduction}
% The very first letter is a 2 line initial drop letter followed
% by the rest of the first word in caps.
%
% form to use if the first word consists of a single letter:
% \IEEEPARstart{A}{demo} file is ....
%
% form to use if you need the single drop letter followed by
% normal text (unknown if ever used by IEEE):
% \IEEEPARstart{A}{}demo file is ....
%
% Some journals put the first two words in caps:
% \IEEEPARstart{T}{his demo} file is ....
%
% Here we have the typical use of a "T" for an initial drop letter
% and "HIS" in caps to complete the first word.

\IEEEPARstart{F}{og}-computing, also known as mobile edge computing (MEC), is a novel and promising technology for future networks \cite{bonomi_fog_2012}. In contrast to conventional base station (BS) and access point (AP), Fog-AP is capable working as a wireless AP, and provides caching and computing capabilities \cite{chiang_fog_2016}. The network composed by Fog-APs is generally referred to as the fog radio access network (Fog-RAN), which can improve the content delivery efficiency and support computation offloading \cite{shi_large-scale_2015}. Therefore, the problem of caching scheme design in Fog-RAN attracted much attention from both industry and academia.

By pushing contents to the edge, the users can access the interested information within one hop, which significantly reduces the latency. In \cite{maddah-ali_fundamental_2014}, an information-theoretic formulation of the caching problem is introduced. The basic structure and global caching gain is analyzed for a novel coded caching approach. In \cite{wang_multi-hop_2018}, the authors focused on content sharing among smart devices in the social IoT with D2D-based cooperative coded caching. In \cite{shanmugam_femtocaching:_2013}, the un-coded and coded caching schemes are studied for femtocells. The fountain code is considered, where the objective is to solve the cache assignment problem with a given maximum number of helpers a user can be connected to. In \cite{ji_fundamental_2016}, the MDS code with random caching strategy is considered in D2D networks, where the fundamental limits of caching is obtained. In \cite{wang_cache_2014}, the authors investigated the content delivery network and information centric network solution of 5G mobile communication network. In \cite{liu_cache_2017}, both centralized algorithm and distributed algorithm are proposed for optimal caching placement problem in Fog-RAN. In \cite{bai_caching_2016}, a caching based socially aware D2D communication framework is considered, where a hypergraph framework is summarized.

Previous works illusatrate great potential of edge caching for reducing the delay by pushing the user required content at network edge nodes. In this paper, we focus on the transmission scheme design and coding parameter optimization for coded caching in Fog-RAN \cite{dimakis_network_2010}. As the Fog-AP may encounter device failure, it is required to recover the stored content from other Fog-APs to a new Fog-AP. By using the coded caching scheme, the data can be recovered by transmitting data from other Fog-APs through backhaul link. Thus, the user can still access the interested content in one hop. On the other hand, the wireless channel may be in deep fading, some of the Fog-APs will be in outage in this case. By using coded caching, the user can recover the interested content by accessing any given number of Fog-APs. In this context, the reliability can be greatly improved if the coded caching scheme is applied. In this work, the objective is to choose the optimal subset of Fog-APs for each user, so that the required content can be deliveried to each user with the highest successful probability, i.e., the content outage probability is minimized. Therefore, we need to choose a coding scheme to minimize the size of data needed to be transmitted from Fog-APs to each user, i.e., the coding scheme with minimum storage size. On the other hand, the Fog-APs are assumed to be connected by reliable (wired or wireless) back-haul links with high bandwidth. The bandwidth requirement for regenerating the storing data in a new Fog-AP can be easily guaranteed. In this context, we choose the minimum storage regenerating (MSR) code with the optimal coding parameters. The transmission scheme design and coding parameter optimization problem is a complicated \emph{multi-objective probabilistic combinatorial optimization} \cite{murat_probabilistic_2006}. Moreover, the requirement on solving this problem in a distributed manner further increases the difficulties.

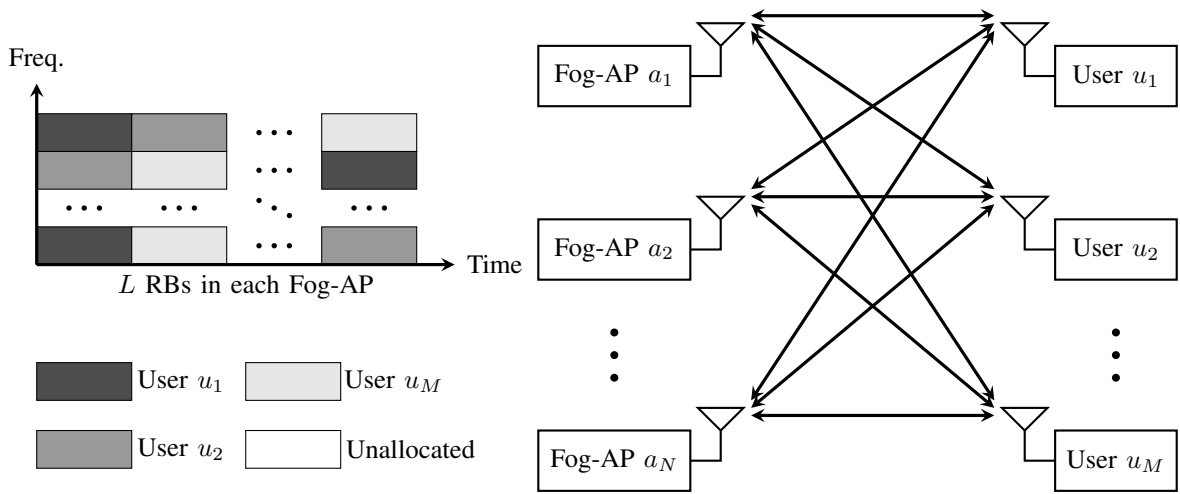
\begin{figure*}[t]
	\centering
	\begin{tikzpicture}[>=stealth]
	    \draw[thick] (-0.4,5.1) rectangle (1.6,5.9);
		\node at (0.6,5.5) {\small Fog-AP $a_1$};
		\draw[thick] (1.6,3.2) -- (2,3.2) -- (2,3.6);
		\draw[thick] (2,3.6) -- (2.3,3.9) -- (1.7,3.9) -- (2,3.6);
	
		\draw[thick] (-0.4,2.8) rectangle (1.6,3.6);
		\node at (0.6,3.2) {\small Fog-AP $a_2$};
		\draw[thick] (1.6,5.5) -- (2,5.5) -- (2,5.9);
		\draw[thick] (2,5.9) -- (2.3,6.2) -- (1.7,6.2) -- (2,5.9);
		
		\fill
		(0.6,1.5) circle (1.5pt)
		(0.6,1.8) circle (1.5pt)
		(0.6,2.1) circle (1.5pt);
		
		\draw[thick] (-0.4,0) rectangle (1.6,0.8);
		\node at (0.6,0.4) {\small Fog-AP $a_N$};
		\draw[thick] (1.6,0.4) -- (2,0.4) -- (2,0.8);
		\draw[thick] (2,0.8) -- (2.3,1.1) -- (1.7,1.1) -- (2,0.8);
	
		\draw[thick] (8,0) rectangle (6.4,0.8);
		\node at (7.2,0.4) {\small User $u_M$};
		\draw[thick] (6.4,0.4) -- (6,0.4) -- (6,0.8);
		\draw[thick] (6,0.8) -- (6.3,1.1) -- (5.7,1.1) -- (6,0.8);
	
		\fill
		(7.2,1.5) circle (1.5pt)
		(7.2,1.8) circle (1.5pt)
		(7.2,2.1) circle (1.5pt);
	
		\draw[thick] (8,2.8) rectangle (6.4,3.6);
		\node at (7.2,3.2) {\small User $u_2$};
		\draw[thick] (6.4,3.2) -- (6,3.2) -- (6,3.6);
		\draw[thick] (6,3.6) -- (6.3,3.9) -- (5.7,3.9) -- (6,3.6);
	
		\draw[thick] (8,5.1) rectangle (6.4,5.9);
		\node at (7.2,5.5) {\small User $u_1$};
		\draw[thick] (6.4,5.5) -- (6,5.5) -- (6,5.9);
		\draw[thick] (6,5.9) -- (6.3,6.2) -- (5.7,6.2) -- (6,5.9);
	
	    \draw[<->,very thick] (2.4,6.1) -- (5.6,1.2);
		\draw[<->,very thick] (2.4,6.2) -- (5.6,4);
		\draw[<->,very thick] (2.4,6.3) -- (5.6,6.3);
		
		\draw[<->,very thick] (2.4,3.8) -- (5.6,1.1);
		\draw[<->,very thick] (2.4,3.9) -- (5.6,3.9);
		\draw[<->,very thick] (2.4,4) -- (5.6,6.2);
		
		\draw[<->,very thick] (2.4,1) -- (5.6,1);
		\draw[<->,very thick] (2.4,1.1) -- (5.6,3.8);
		\draw[<->,very thick] (2.4,1.2) -- (5.6,6.1);
	
		\filldraw[fill=black!70] (-7,3) rectangle (-5.75,3.5);
		\filldraw[fill=black!10] (-5.75,3) rectangle (-4.5,3.5);
		\filldraw[fill=black!40] (-3.25,3) rectangle (-2,3.5);
		
		\fill
		(-6.575,3.75) circle (1pt)
		(-6.375,3.75) circle (1pt)
		(-6.175,3.75) circle (1pt)
		(-5.325,3.75) circle (1pt)
		(-5.125,3.75) circle (1pt)
		(-4.925,3.75) circle (1pt)
		(-4.075,3.85) circle (1pt)
		(-3.875,3.75) circle (1pt)
		(-3.675,3.65) circle (1pt)
		(-2.825,3.75) circle (1pt)
		(-2.625,3.75) circle (1pt)
		(-2.425,3.75) circle (1pt);
		
		\filldraw[fill=black!40] (-7,4) rectangle (-5.75,4.5);
		\filldraw[fill=black!10] (-5.75,4) rectangle (-4.5,4.5);
		\filldraw[fill=black!70] (-3.25,4) rectangle (-2,4.5);
		
		\filldraw[fill=black!70] (-7,4.5) rectangle (-5.75,5);
		\filldraw[fill=black!40] (-5.75,4.5) rectangle (-4.5,5);
		\filldraw[fill=black!10] (-3.25,4.5) rectangle (-2,5);
	
	    \fill
		(-4.075,3.25) circle (1pt)
		(-3.875,3.25) circle (1pt)
		(-3.675,3.25) circle (1pt)
		(-4.075,4.25) circle (1pt)
		(-3.875,4.25) circle (1pt)
		(-3.675,4.25) circle (1pt)
		(-4.075,4.75) circle (1pt)
		(-3.875,4.75) circle (1pt)
		(-3.675,4.75) circle (1pt);
		
		\draw[->,very thick] (-7,3) -- (-1.5,3) node[right] {\small Time};
		\draw[->,very thick] (-7,3) -- (-7,5.4) node[above] {\small Freq.};
		\node at (-4.25,2.7) {\small $L$ RBs in each Fog-AP};
	
		\filldraw[fill=black!70] (-7,1.2) rectangle (-5.75,1.7);
		\node at (-5.1,1.45) {\small User $u_1$};
		\filldraw[fill=black!40] (-7,0.3) rectangle (-5.75,0.8);
		\node at (-5.1,0.55) {\small User $u_2$};
		\filldraw[fill=black!10] (-4.25,1.2) rectangle (-3,1.7);
		\node at (-2.3,1.45) {\small User $u_M$};
		\draw (-4.25,0.3) rectangle (-3,0.8);
		\node at (-2.08,0.55) {\small Unallocated};
	\end{tikzpicture}
	\caption{Coded caching in Fog-RAN, where $M$ contents are stored at $N$ Fog-APs with MSR code. Each users will access $K$ Fog-APs to fetch the interested content. Each Fog-AP has $L$ RBs, each of which will be allocated to at most one user. The content outage probability defined in \eqref{eq:outage_definition} is used as the performance metric for each user.}\label{fig:system_desc}
\end{figure*}

It is well known there is (generally) no global optimum to multi-objective optimization problem \cite{bjornson_multiobjective_2014}. Therefore, we focus on the Pareto optimal solution (i.e., none of the objectives can be improved without degrading other objectives) with fairness constraint for coded caching problem in Fog-RAN. The key is the novel $b$-matching approach, which can solve this problem with fixed coding parameters. Here $b$ is a function defined on vertices, which determines the maximum number of edges associated with a vertex in the matching \cite{huang_fast_2011}. Specifically, the Fog-AP selection problem is first formulated as a random bipartite graph (RBG) based fairness maximum $b$-matching problem. To explore not only the caching capability but also computing capability of Fog-APs, a fast message passing algorithm is proposed to select Fog-APs for each user in a distributed way. The memory usage of the proposed algorithm is only $O(M+N)$, where $M$ is the number of users and $N$ is the number of Fog-APs. Although, it is very difficult to derive the closed-form formulas for the optimal solution, a tight upper bound for the solution is still obtained by digging the properties of fairness maximum $b$-matching on RBG and conditional content outage probability. To obtain the optimal coding parameters, we define and derive the outage exponent region (OER), the region that contains all feasible outage exponent vectors of all users, and diversity-multiplexing region (DMR), the OER as the signal to noise ratio (SNR) tends to infinity, from content outage probability.\footnote{The diversity-multiplexing tradeoff (DMT), which is first proposed in \cite{zheng_diversity-multiplexing_2002}, is used as a key performance metric in high SNR regime. Specifically, the DMT is the slope of the outage probability curve in the high SNR regime. According to the definition in \cite{zheng_diversity-multiplexing_2002}, the DMT is the tradeoff between reliability (diversity gain) and efficiency (multiplexing gain) when SNR tends to infinity. The DMR is the DMT in multiple user scenario. The OER is a generalization of DMR in finite SNRs with multiple users.} The obtained theoretical results illustrate that $b$-matching approach is optimal in high SNR regime. Specifically, each user not only fairly shares the multiplexing gain but also achieves the full diversity gain, i.e.,the largest area of DMR. Simulation results illustrate the accuracy of the theoretical derivations, and verify the outage performance of the proposed scheme. In general, the contribution our work can be summarized as follows:
\begin{enumerate}
	\item The optimal content transmission scheme from Fog-APs and users is proposed, which is formulated as a multi-objective probabilistic combinatorial optimization problem. The novel $b$-matching approach is proposed for solving this problem. The decentralized algorithm is also obtained to select the optimal subset of Fog-APs for each user with low complexity and small storage cost.
	\item The closed-form approximation formulas of the content outage probability is derived in this work. The DMR can be obtained to verify that the solution is optimal in high SNRs. These formulas provide a systematic way to design and evaluate the caching scheme for Fog-RANs in an analytic way.
\end{enumerate}

The rest of this paper is organized as follows. Section \ref{sec:system_model} presents the system model and problem formulation. The RBG based $b$-matching approach is presented in Section \ref{sec:f_matching}, where the fast message passing based distributed Fog-AP selection algorithm is discussed in detail. Section \ref{sec:outage_analysis} defines and derives the closed-form formulas for content outage probabilities, OER, and DMR to obtain the optimal coding parameters. Section \ref{sec:simulation_results} presents the simulation results. Finally, Section \ref{sec:conclusions} concludes this paper.

\section{System Model and Problem Formulation}\label{sec:system_model}

\subsection{Channel Model}\label{subsec:channel_model}

Consider a dense Fog-RAN, as shown in Fig. \ref{fig:system_desc}, where $M$ users communicate with $N$ Fog-APs through orthogonal subchannels. The user set is denoted by $\mathcal{U}=\{u_m\}_{m=1}^M$, where $u_m$ is the $m$-th user.\footnote{The script symbol $\mathcal{X}$ denotes a set, whose cardinality is denoted by $|\mathcal{X}|$. The abbreviation $\{x_m\}_{m=1}^M$ is used to denote $\{x_1,x_2,\ldots,x_M\}$.} The Fog-AP set is denoted by $\mathcal{A}=\{a_n\}_{n=1}^N$, where $a_n$ is the $n$-th Fog-AP. One Fog-AP has a dedicated subchannel, each of which contains $L$ resource blocks (RBs) in the duration of the coherence time. For convenience, both the bandwidth and the duration of one RB are normalized as one. In this context, one Fog-AP can be accessed by $L$ users at most, each of which occupies one RB of the subchannel.

Suppose that the signal on each subchannel undergoes the independent Rayleigh slow fading. Then, the channel gain between user $u_m$ and Fog-AP $a_n$, denoted by $h_{mn}$, is independent and follows the identical distribution of $\mathcal{CN}(0,1)$.\footnote{$\mathcal{CN}(\mu,\sigma^2)$ denotes a circularly symmetric Gaussian distribution with mean $\mu$ and variance $\sigma^2$.} The received signal from Fog-AP $a_n$ to user $u_m$, denoted by $y_m$, is given by
\begin{equation*}
	y_m=h_{mn}x_n+w_m,
\end{equation*}
where $w_m\sim\mathcal{CN}(0,1)$ is the additive Gaussain white noise (AWGN). As we assume one bit channel state information (CSI) is known at the transmitter, the mutual information between user $u_m$ and Fog-AP $a_n$ during one RB is given by
\begin{equation}\label{eq:subchannel_capacity}
	I_{mn}=\max I(y_m;x_n)=\ln\left(1+|h_{mn}|^2\gamma\right)\quad\unit{nats},
\end{equation}
where $\gamma$ is the average SNR at the receiver. Throughout the paper, the unit of information is ``$\unit{nat}$''. As we normalize both the bandwidth and the duration of one RB as one, the transmission rate is equal to $\unit[I_{mn}]{nats}$ per RB use.

\subsection{Coded Caching in Fog-RANs}\label{subsec:CodedCaching}

To fully exploit the caching capability in Fog-APs, the coded caching scheme is adopted. Suppose the user $u_m$ requires the $m$-th content $s_m$, which composes the content set $\mathcal{S}=\{s_m\}_{m=1}^M$. The size of content $s_m$ is $\unit[R_m]{nats}$. Thus, there are $M$ files cached in Fog-RAN. As proposed in \cite{dimakis_network_2010}, the distributed storage code with parameters $(N,K_m,D_m)$ is used for content $s_m$. Thus, $s_m$ will be cached in $N$ Fog-APs, each of which stores $\unit[\alpha_m]{nats}$. Hence, the storage size of each Fog-AP is at least $\unit[\sum_{m=1}^M\alpha_m]{nats}$. If $K_m$ Fog-APs transmit $\unit[\alpha_m]{nats}$ each to user $u_m$, it is capable of recovering the original content $s_m$. For a new Fog-AP without caching any part of $s_m$, it requires $D_m$ Fog-APs to transmit $\unit[\beta_m]{nats}$ in total to regenerate $\unit[\alpha_m]{nats}$ in this new Fog-AP. It can be seen that the performance of the distributed storage code is characterized by $(\alpha_m,\beta_m)$ \cite{dimakis_network_2010}. In this context, we have $N\geq K_m$ for $m=1,2,\ldots,M$ and $\sum_{m=1}^MK_m\leq NL$.

In the considered Fog-RAN, the contents should be first cached in Fog-APs based on some caching strategies \cite{liu_cache_2017,maddah-ali_fundamental_2014}. Then, the method in this work will be applied to optimize the transmission performance for the cached contents. Specifically, Fog-APs are connected by reliable (wired or wireless) back-haul links with high bandwidth \cite{shi_large-scale_2015}. Thus, compared to the wireless channel between users and Fog-APs, the total regenerating bandwidth $\beta_m$ can be naturally assumed to be fulfilled.\footnote{If we take back-haul links into consideration, there will be a tradeoff between $\alpha_m$ and $\beta_m$. The proposed framework in this paper can be easily extended to this scenario.} In contrast, due to channel fading, the mutual information between user $u_m$ and Fog-AP $a_n$ during one RB may be smaller than $\alpha_m$. In other words, the Fog-AP $a_n$ is in outage for user $u_m$, referred to as \emph{Fog-AP outage}. Thus, we need to minimize $\alpha_m$, i.e., apply MSR codes \cite{dimakis_network_2010}. Accordingly, we have the following optimal $(\alpha_m^*,\beta_m^*)$ for MSR codes:
\begin{equation}\label{eq:MSRcode}
	(\alpha_m^*,\beta_m^*)=\left(\frac{R_m}{K_m},\frac{D_m}{D_m-K_m+1}\frac{R_m}{K_m}\right).
\end{equation}

\subsection{Fog-AP Selection Problem}

According to the property of MSR codes, the user must successfully access $K_m$ Fog-APs in order to recover the required content $s_m$. Due to channel fading, however, the Fog-AP $a_n$ will be in outage for user $u_m$, if $I_{mn}<\alpha_m^*$. In this context, the user may still fail to recover $s_m$, even the MSR code with $(\alpha_m^*,\beta_m^*)$ is applied, i.e., the user may encounter \emph{content outage}. Intuitively, with the fixed transmission rate $\alpha_m^*$ from each Fog-AP to user $u_m$ in one RB, the content outage probability decreases if we decrease $K_m$. However, as $\alpha_m^*$ is equal to $\frac{R_m}{K_m}$, the Fog-AP outage probability increases if we decrease $K_m$. Therefore, an optimal $K_m$ exists for achieving the minimum content outage probability. On the other hand, because of the flexible architecture of the Fog-RAN, several Fog-APs are able to form an edge cloud to support mobile edge computing. Thus, the joint channel coding scheme, such as rotated $\mathbb{Z}^{K_m}$-lattice code \cite{oggier_cyclic_2007} or permutation code \cite{tse_fundamentals_2005}, can be applied in Fog-APs so as to further improve the outage performance. According to \cite{oggier_cyclic_2007,tse_fundamentals_2005,bai_outage_2013}, the outage probability of user $u_m$ for content $s_m$ is given by
\begin{equation}\label{eq:outage_definition}
	p_m^\mathrm{out}(R_m)=\Pr\left\{\sum_{a_n\in\mathcal{A}_m}I_{mn}<R_m\right\},
\end{equation}
where $R_m=K_m\alpha_m^*$, and $\mathcal{A}_m$ is the set of Fog-APs accessable by user $u_m$, which satisfies
\begin{equation}
	\left\{
	\begin{aligned}
		& |\mathcal{A}_m|=K_m; \\
		& \bigcup_{m=1}^M\mathcal{A}_m\subseteq\mathcal{A}.
	\end{aligned}
	\right.
\end{equation}

To reduce the complexity in both signaling and computation, each user is only allowed to feedback $\unit[1]{bit}$ CSI for each subchannel to Fog-APs at the beginning of each coherence time, i.e., we only knows whether the Fog-AP $a_n$ is in outage for user $u_m$ or not. Let $\mathbf{Q}$ denote the $\unit[1]{bit}$ quantized CSI matrix, the entry at the $m$-th row and $n$-th column is given by
\begin{equation}\label{Eq: quantized CSI matrix}
	[\mathbf{Q}]_{mn}=\left\{
	\begin{aligned}
		& 0, && I_{mn}<\alpha_m^*; \\
		& 1, && I_{mn}\geq\alpha_m^*.
	\end{aligned}
	\right.
\end{equation}
Therefore, the Fog-AP outage probability is given by
\begin{equation}\label{eq:subchannel_outage}	
	p_m=\Pr\{[\mathbf{Q}]_{mn}=0\} = 1-\exp\left(-\frac{e^{\alpha_m^*}-1}{\gamma}\right).
\end{equation}
For convenience, we define
\begin{equation}	
	q_m=\Pr\{[\mathbf{Q}]_{mn}=1\} = 1-p_m = \exp\left(-\frac{e^{\alpha_m^*}-1}{\gamma}\right).
\end{equation}
In this context, a Fog-AP selection scheme can be seen as a mapping from $\mathbf{Q}$ to $\{\mathcal{A}_m\}_{m=1}^M$, that is
\begin{equation}
	\mathscr{S}:\mathbf{Q}\to\{\mathcal{A}_m\}_{m=1}^M.
\end{equation}
According to the notation in \cite{bjornson_multiobjective_2014} and footnote 1 of this paper, the optimal Fog-AP selection problem can be formulated as follows:
\begin{equation}\label{eq:MOPCO}
	\begin{aligned}
		\text{(P1)}\quad\min_\mathscr{S} & && \left\{p_m^\mathrm{out}(R_m)\right\}_{m=1}^M \\
		\mathrm{s.t.} & && \sum_{m=1}^M K_m\leq NL \\
					  & && K_m<N.
	\end{aligned}
\end{equation}
The problem P1 in Eq. \eqref{eq:MOPCO} summarizes the optimal Fog-AP selection problem in an abstraction form. It will determine the optimal mapping $\mathscr{S}$ and $\{K_m\}_{m=1}^M$ so as to minimize the content outage probability in Eq. \eqref{eq:outage_definition}. As a \emph{multi-objective probabilistic combinatorial optimization} problem, P1 has no global optimum in general according to \cite{murat_probabilistic_2006,bjornson_multiobjective_2014}. Therefore, the Pareto optimal is proposed as a solution for multi-objective optimization problem. Specifically, the strong Pareto boundary of a multi-objective optimization problem consists of the attainable operating points that none of the objectives can be improved without degrading other objectives. Every point on the strong Pareto boundary is a Pareto optimal solution. In this work, we focus on the Pareto optimal solution with fairness consideration, denoted by $\mathscr{S}^*$ and $\{K_m^*\}_{m=1}^M$ \cite{bjornson_multiobjective_2014}.\footnote{According to \cite{bjornson_multiobjective_2014}, the scalarization of this problem can be obtained by specifying a certain subjective tradeoff between the objectives. In this work, the Pareto optimal with fairness constraint is used as the scalarization method. As an another scalarization of the problem, the min-max formulation will be investigated in our future work.} In the following, with fixed $K_m$ P1 is first reformulated and solved as P2 by the proposed $b$-mathcing approach in Section \ref{sec:f_matching}. The $b$-matching approach guarantees that all of the users fairly share the total channel resource. Specifically, each user will be allocated the channel resource which is proportional to its rate requirement. In Section \ref{sec:outage_analysis}, the content outage probability, as a function of $K_m$, is obtained in closed-form formulas, so that $K_m$ can be optimized. The theoretical results show that the $b$-matching approach also achieves the optimal diversity order, i.e., the total diversity gain provided by the channel. Therefore, the Pareto optimal solution with fairness constraint is obtained by the proposed $b$-matching approach.

\section{RBG based $b$-Matching Approach}\label{sec:f_matching}

In this section, P1 with fixed $K_m$ is first reformulated as a combinatorial optimization problem P2 based on a given sample of $\mathbf{Q}$. Then, the RBG based $b$-matching method is proposed for solving P2, which is exactly the Pareto optimal mapping $\mathscr{S}^*$ with fairness consideration for P1. To fully exploit the computation capability of Fog-APs, a fast message passing based distributed Fog-AP selection algorithm is proposed.

\subsection{Problem Re-formulation}\label{subsec:RBG_Formulation}

At the beginning epoch of each coherence time, $\mathbf{Q}$ can be obtained by channel estimation. The optimal Fog-AP selection problem will then be solved based on the known $\mathbf{Q}$. Define the decision variable $x_{mn}$ as
\begin{equation}
	x_{mn}=\left\{
	\begin{aligned}
		& 1, && \text{if }a_n\in\mathcal{A}_m; \\
		& 0, && \text{otherwise}.
	\end{aligned}
	\right.
\end{equation}
The decision matrix with $x_{mn}$ being the entry at the $m$-th row and $n$-th column is denoted by $\mathbf{X}$. With the fixed $K_m$ and known $\mathbf{Q}$, the mapping $\mathscr{S}$ can be represented by the decision matrix $\mathbf{X}$. To achieve the Pareto optimal solution, i.e., minimum content outage probability, the objective is to maximize the total number of non-outage Fog-APs which are accessed by all of the users. The fairness constraint means that $u_m$ will access $\eta K_m$ non-outage Fog-APs has the same probability $\epsilon$, where $\eta<1$ and $\epsilon<1$. In this context, the fair Pareto optimal mapping $\mathscr{S}^*$ is equivalent to solve the following combinatorial optimization problem:
\begin{equation}\label{eq:b_matching_problem}
	(\text{P2})\quad
	\begin{aligned}
		\max_\mathbf{X} & && \sum_{m=1}^M\sum_{n=1}^N[\mathbf{Q}]_{mn}x_{mn} \\
		\mathrm{s.t.}   & && \sum_{m=1}^M x_{mn}\leq L, \\
					    & && \sum_{n=1}^N x_{mn}=K_m, \\
					    & && \Pr\left\{\frac{\sum_{n=1}^N[\mathbf{Q}]_{mn}x_{mn}}{K_m}=\eta\right\}=\epsilon, \\
					    & && x_{mn}\in\{0,1\}, \\
					    & && m=1,\ldots,M,\quad n=1,\ldots,N.
	\end{aligned}
\end{equation}
It can be seen that the content outage probability in P1 is minimized when combining the optimal solution of P2 and joint coding scheme for every sample of $\mathbf{Q}$. The first constraint implies that each Fog-AP can only support $L$ users, each of which occupies one RB. The second constraint indicates that each user will access $K_m$ Fog-APs to fulfill the requirement of MSR codes. Due to channel fading, however, the accessible $K_m$ Fog-APs for user $u_m$ may not always be non-outage for every sample of $\mathbf{Q}$. Therefore, the third constraint is introduced to guarantee fairness.

\subsection{$b$-Matching based Fog-AP Selection}\label{sec:max_b_matching}

P2 is a complicated combinatorial optimization problem. The RBG based $b$-matching method is proposed in the following to solve P2, which is also the Pareto optimal mapping $\mathscr{S}^*$ with fairness consideration for P1. The preliminaries on RBG and $b$-matching can be found in Appendix \ref{app:RBG_b_matching}.

\begin{figure}[t]
	\centering
	\begin{tikzpicture}
		\fill
		(-2,3) circle (3pt)
		(-2,2) circle (3pt)
		(-2,1) circle (3pt)
		(-2,0) circle (3pt);
		
		\node at (-2.5,3) {$u_1$};
		\node at (-2.5,2) {$u_2$};
		\node at (-2.5,1) {$u_3$};
		\node at (-2.5,0) {$u_4$};
		
		\draw[line width=0.5pt,style=dashed] (-2.8,-0.5) rectangle (-1.7,3.5);
		\node at (-2.25,3.8) {Users};
				
		\fill
		(2,2.5) circle (3pt)
		(2,1.5) circle (3pt)
		(2,0.5) circle (3pt);		

		\node at (2.5,2.5) {$a_1$};
		\node at (2.5,1.5) {$a_2$};
		\node at (2.5,0.5) {$a_3$};
		
		\draw[line width=0.5pt,style=dashed] (1.7,-0.5) rectangle (2.8,3.5);
		\node at (2.25,3.8) {Fog-APs};
		
		\draw [line width=2pt]
		(-2,3) -- (2,2.5)
		(-2,3) -- (2,0.5)
		(-2,2) -- (2,2.5)
		(-2,1) -- (2,1.5)
		(-2,1) -- (2,0.5)
		(-2,0) -- (2,1.5)
		(-2,0) -- (2,0.5);
		
		\draw [line width=2pt,dashed] (-2,2) -- (2,1.5);
	\end{tikzpicture}
	\caption{A sample of $\mathscr{G}\{\mathcal{K}_{4,3};\mathsf{P}\}$ with $K_m=2$ and $L=3$. The thick segments are the edges in $\mathcal{M}_b^\mathrm{fm}$. The dashed segment denotes the selected outage Fog-AP.}\label{fig:RBG_example}
\end{figure}
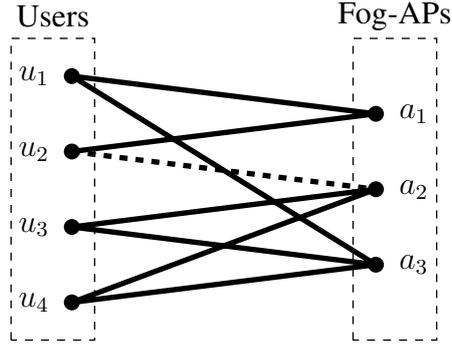

The RBG model for Fog-RAN can be constructed as follows. All of the vertices in one partition class represent all of the users in $\mathcal{U}$, and all of the vertices in the other partition class represent all of the Fog-APs in $\mathcal{A}$. If $[\mathbf{Q}]_{mn}=1$, we join the vertex $u_m\in\mathcal{U}$ and the vertex $a_n\in\mathcal{A}$ with an edge $e=(u_m,a_n)$. Otherwise, there will be no edge between $u_m$ and $a_n$. Therefore, the probability space of the RBG model can be denoted by $\mathscr{G}\{\mathcal{K}_{MN};\mathsf{P}\}$ with $\mathsf{P}=\{p_m\}_{m=1}^M$, where $p_m$ is given by Eq. \eqref{eq:subchannel_outage}. A sample of $\mathscr{G}\left\{\mathcal{K}_{4,3};\mathsf{P}\right\}$ is shown in Fig. \ref{fig:RBG_example}. 

It can be seen that each sample of $\mathscr{G}\{\mathcal{K}_{MN};\mathsf{P}\}$, denoted by a bipartite graph $\mathcal{G}(\mathcal{U}\cup\mathcal{A},\mathcal{E})$, corresponds to a snapshot of Fog-RAN and vice versa. For any given $\mathcal{G}(\mathcal{U}\cup\mathcal{A},\mathcal{E})$, associate $b(v)$ to $v\in\mathcal{U}\cup\mathcal{A}$ as
\begin{equation}\label{eq:b_v}
	b(v)=\left\{
	\begin{aligned}
		& K_m,	&& v=u_m\in\mathcal{U}; \\
		& L,	&& v=a_n\in\mathcal{A}.
	\end{aligned}
	\right.
\end{equation}
As the maximum $b$-matching $\mathcal{M}_b^\mathrm{m}$ may not be unique on some samples, to guarantee the third constraint in problem P2, we need to generate $\mathcal{M}_b^\mathrm{m}$ on $\mathcal{G}(\mathcal{U}\cup\mathcal{A},\mathcal{E})$ with $\Pr\left\{\frac{k_m}{K_m}=\eta\right\}=\epsilon$ for every $u_m\in\mathcal{U}$, where $k_m=|\mathcal{M}_b^\mathrm{m}\cap\mathcal{E}(u_m)|$ is the number of non-outage Fog-APs allocated to $u_m$. The maximum $b$-matching satisfying the fairness constraint is referred to as the \emph{fairness maximum $b$-matching} and denoted by $\mathcal{M}_b^\mathrm{fm}$. In this context, the Fog-AP set $\mathcal{A}_m^*$ selected for user $u_m$ can be constructed as
\begin{equation}
	\mathcal{A}_m^*=\mathcal{A}(\mathcal{M}_b^\mathrm{fm}\cap\mathcal{E}(u_m))\cup\tilde{\mathcal{A}}_m,
\end{equation}
where $\mathcal{A}(\mathcal{M}_b^\mathrm{fm}\cap\mathcal{E}(u_m))\subseteq\mathcal{A}$ contains all of the vertices that are incident with an edge in $\mathcal{M}_b\cap\mathcal{E}(u_m)$. $\tilde{\mathcal{A}}_m\subseteq\mathcal{A}$ contains $K_m-k_m$ Fog-APs, each of which satisfies 
\begin{equation}
	\sum_{m'=1}^M\mathbf{1}_{\mathcal{A}_{m'}^*}(a_n)\leq L,\quad \forall a_n\in\tilde{\mathcal{A}}_m,
\end{equation}
where $\mathbf{1}_{\mathcal{X}}(x)$ is the indicator function defined as
\begin{equation}
	\mathbf{1}_{\mathcal{X}}(x)=
	\left\{
	\begin{aligned}
		& 1, && x\in\mathcal{X}; \\
		& 0, && x\notin\mathcal{X}.
	\end{aligned}
	\right.
\end{equation}
Clearly, if $\mathcal{M}_b^\mathrm{fm}=\mathcal{M}^\mathrm{p}_b$, $\tilde{\mathcal{A}}_m$ is an empty set for any $u_m\in\mathcal{U}$. Otherwise, $\tilde{\mathcal{A}}_m$ can be generated by randomly selecting $K_m-k_m$ non-saturated outage Fog-APs to user $u_m$. The discussion above implies that the optimal solution of P2 is given by:
\begin{equation}\label{eq:OptimalP2}
	x_{mn}^*=\left\{
		\begin{aligned}
			& 1, && \text{if }a_n\in\mathcal{A}_m^*; \\
			& 0, && \text{otherwise}.
		\end{aligned}
		\right.
\end{equation}

In the case of $\tilde{\mathcal{A}}_m\neq\emptyset$, the content outage probability for user $u_m$ in P1 also depends on the applied joint coding scheme. Therefore, after generating the optimal solution of P2 as Eq. \eqref{eq:OptimalP2}, the optimal joint coding scheme will then be applied on $\mathcal{A}_m^*$ for user $u_m$. An example of $\mathcal{M}_b^\mathrm{fm}$ is shown in Fig. \ref{fig:RBG_example}. It can be seen that $u_1$, $u_3$ and $u_4$ are $\mathcal{M}_b^\mathrm{fm}$-saturated, and thus are not in outage. The outage state of $u_2$, however, is determined by the sum capacity achieved by the applied coding scheme.

\subsection{Distributed Fog-AP Selection Algorithm}

Based on the fast message passing method in \cite{pmlr-v15-huang11a}, a distributed Fog-AP selection algorithm is proposed to solve problem P2 in this subsection. As discussed in Section \ref{sec:max_b_matching}, the optimal Fog-AP selection can be obtained by finding the fairness maximum $b$-matching $\mathcal{M}_b^\mathrm{fm}$. By representing the maximum $b$-matching problem in a factor graph, it can be mapped into a marginal probability computation problem on the probabilistic graphical model. Therefore, the maximum $b$-matching problem can then be solved by passing messages between the adjacent vertices on the probabilistic graphical model. The fairness is guaranteed by adding randomness in the message passing process. The basic message passing algorithm for solving the fairness maximum $b$-matching problem is similar to the algorithm in \cite{Bai_TWC}, whose total computation cost scales as $O(NM^2)$.

To reduce the memory usage to $O(M+N)$, an improved message passing algorithm proposed in \cite{pmlr-v15-huang11a} is applied to solve the fairness maximum $b$-matching problem. A key step of this algorithm is the selection operation, which finds the $k$-th largest element in a set $\mathcal{X}$ for a given $k$. The selection operation over $\mathcal{X}$ can be written as
\begin{equation}\label{eq:selection_operation}
    \sigma_k(\mathcal{X})=\{x\in\mathcal{X}:|\{t\in\mathcal{X}:t\geq x\}|=k\}.
\end{equation}
For convenience, we define
\begin{equation}
    \tilde{\mathbf{Q}}=\left[
    \begin{array}{cc}
        \mathbf{O} & \mathbf{Q} \\
        \mathbf{Q}^T & \mathbf{O} 
    \end{array}
    \right],
\end{equation}
where $\mathbf{Q}$ is the $\unit[1]{bit}$ quantized CSI matrix, and
\begin{equation}
    \tilde{\mathbf{X}}=\left[
    \begin{array}{cc}
        \mathbf{O} & \mathbf{X} \\
        \mathbf{X}^T & \mathbf{O} 
    \end{array}
    \right],
\end{equation}
where $\mathbf{X}$ is the decision matrix in problem P2. The message passing algorithm maintains a belief value for each edge in $\mathcal{G}(\mathcal{V}\cup\mathcal{A},\mathcal{E})$, which can be denoted by a matrix $\mathbf{B}$. Similarly, we define
\begin{equation}
    \tilde{\mathbf{B}}=\left[
    \begin{array}{cc}
        \mathbf{O} & \mathbf{B} \\
        \mathbf{B}^T & \mathbf{O} 
    \end{array}
    \right].
\end{equation}

The belief value for the edge $e_{ij}=(v_i,v_j)$ at iteration $t$ is denoted by the entry $[\tilde{\mathbf{B}}]_{ij}^t$, where $v_i=u_m$ for $i=1,2,\ldots,M$ and $v_i=a_n$ for $i=M+1,M+2,\ldots,M+N$. $[\tilde{\mathbf{B}}]_{ij}^t$ is updated with the following rule
\begin{equation}\label{eq:belief}
    [\tilde{\mathbf{B}}]_{ij}^t=[\tilde{\mathbf{Q}}]_{ij}-\sigma_L\left(\left\{[\tilde{\mathbf{B}}]_{jk}^{t-1}:v_k\in\mathcal{U}\cup\mathcal{A},k\neq i\right\}\right).
\end{equation}
Define $\mu_j$ as the negation of the $b(v_j)$-th selection and $\nu_j$ for the $(b(v_j)+1)$-th selection. $\mu_j$ and $\nu_j$ are updated with the following rules
\begin{equation}\label{eq:parameter}
    \left\{
    \begin{aligned}
        \mu_j^t & =-\sigma_{b(v_j)}\left(\left\{[\tilde{\mathbf{B}}]_{jk}^{t-1}:v_k\in\mathcal{U}\cup\mathcal{A}\right\}\right); \\
        \nu_j^t & =-\sigma_{b(v_j)+1}\left(\left\{[\tilde{\mathbf{B}}]_{jk}^{t-1}:v_k\in\mathcal{U}\cup\mathcal{A}\right\}\right).
    \end{aligned}
    \right.
\end{equation}
Let $\tilde{x}_{ij}^t$ denote the entry at the $i$-th row and $j$-th column of $\tilde{\mathbf{X}}$ at the $t$-th iteration. The resulting belief searching rule can then be written as
\begin{equation}\label{eq:belief_lookup_rule}
    [\tilde{\mathbf{B}}]_{ij}^t=\left\{
	\begin{aligned}
		& [\tilde{\mathbf{Q}}]_{ij}+\mu_j^t, && \tilde{x}_{ji}^t=1; \\
		& [\tilde{\mathbf{Q}}]_{ij}+\nu_j^t, && \text{otherwise}.
	\end{aligned}
	\right.
\end{equation}
The estimation of $\tilde{\mathbf{X}}$ after each iteration will be updated as
\begin{equation}\label{eq:CurrentEstimate_S}
    \tilde{x}_{mn}^t=\left\{
	\begin{aligned}
		& 1, && \mu_i^t\leq[\tilde{\mathbf{B}}]_{ij}^{t-1}; \\
		& 0, && \text{otherwise}.
	\end{aligned}
	\right.
\end{equation}
It is shown in \cite{pmlr-v15-huang11a} that this algorithm converges if there exists a valid $b$-matching solution. The details of this algorithm are summarized in Algorithm \ref{alg:Al1}.

\begin{algorithm}[t]
\caption{Distributed Fog-AP Selection}
    \begin{algorithmic}[1]\label{alg:Al1}
        \STATE $\mu_j^0, \nu_j^0\leftarrow0$ for $j=1,2,\ldots,M+N$
        \STATE $\tilde{\mathbf{X}}^0\leftarrow\emptyset$
        \STATE $t\leftarrow1$
        \WHILE{not converge}
            \FOR{$j=1,\ldots,M+N$}
                \STATE $\tilde{x}_{jk}^t\leftarrow0$, for $k=1,2,\ldots,M+N$
                \STATE $\mu_j^t\leftarrow-\sigma_{b(j)}\left(\left\{[\tilde{\mathbf{B}}]_{jk}^{t-1}:v_k\in\mathcal{U}\cup\mathcal{A}\right\}\right)$
                \STATE $\nu_j^t\leftarrow-\sigma_{b(j)+1}\left(\left\{[\tilde{\mathbf{B}}]_{jk}^{t-1}:v_k\in\mathcal{U}\cup\mathcal{A}\right\}\right)$
                \FORALL{$\{k:\mu_j^t\leq[\tilde{\mathbf{B}}]_{jk}^{t-1}\}$}
                    \STATE $\tilde{x}_{jk}^t\leftarrow1$
                \ENDFOR
            \ENDFOR
            \STATE Delete $\tilde{\mathbf{X}}^{t-1}$, $\mu_j^{t-1}$, and $\nu_j^{t-1}$
            \STATE $t \leftarrow t+1$
        \ENDWHILE
        \STATE Output $\tilde{\mathbf{X}}$
    \end{algorithmic}
\end{algorithm}

To further reduce the running time, a variation of Algorithm \ref{alg:Al1} is introduced, where Steps 7-8 are improved. Since $\tilde{\mathbf{Q}}$ does not change during the message passing, the algorithm can compute the index cache $\mathbf{I}\in\mathbb{N}^{(M+N)\times c}$ with cache size $c$ initially, where $[\mathbf{I}]_{ik}$ is the index of the $k$-th largest weight connected to node $v_i$. 

For $\kappa=[\mathbf{I}]_{ik}$,
\begin{equation}\label{eq:weight_update}
    [\tilde{\mathbf{Q}}]_{i\kappa}=\sigma_k\left(\left\{[\tilde{\mathbf{Q}}]_{ij}:v_j\in\mathcal{U}\cup\mathcal{A}\right\}\right).
\end{equation}
As for $\theta=d_k$ and entry $\nu_\theta^t=\sigma_k(\nu_j^t:v_j\in\mathcal{U}\cup\mathcal{A})$, the $\nu^t$ values are similarly sorted and stored in the index vector $\mathbf{d}\in\mathbb{N}^{M+N}$ after each iteration. After updating process of each step, we maintain a set $\mathcal{B}$ of the greatest $b(v_j)+1$ beliefs obtained so far. Thus, the current estimation for $\mu_j^t$ and $\nu_j^t$ at each stage are
\begin{equation}\label{eq:parameter_update}
    \left\{
	\begin{aligned}
		\hat{\mu}_j^t & =\sigma_{b(v_j)}(\mathcal{B}),\\
		\hat{\nu}_j^t & =\min(\mathcal{B}).
	\end{aligned}
	\right.
\end{equation}
At each iteration, the greatest possible unseen belief is bounded as the sum of the least weight seen so far from the sorted weight cache and the least $\nu$ value so far from the $\nu$ cache. The sufficient selection procedure will exit when $\hat{\nu}_j^t$ becomes less than or equals to the sum, since further comparisons are unnecessary. The details of this algorithm is shown in Algorithm \ref{alg:Al2}.

\begin{algorithm}[t]
\caption{Sufficient Selection}
    \begin{algorithmic}[1]\label{alg:Al2}
        \STATE $k\leftarrow1$
        \STATE $\rho\leftarrow\infty$
        \STATE $\mathcal{B}\leftarrow\emptyset$
        \STATE $\hat{\mu}_j^t\leftarrow-\infty$
        \STATE $\hat{\nu}_j^t\leftarrow-\infty$
        \WHILE{$\hat{\nu}_j^t<\rho$}
            \IF{$k\leq c$}
                \STATE $\kappa\leftarrow[\mathbf{I}]_{jk}$
                \IF{$\kappa$ is unvisited and $[\tilde{\mathbf{B}}]_{j\kappa}^{t-1}>\min (\mathcal{B})$}
                    \STATE $\mathcal{B}\leftarrow(\mathcal{B}\setminus\min(\mathcal{B}))\cup[\tilde{\mathbf{B}}]_{j\kappa}^{t-1}$
                \ENDIF
            \ENDIF
            \STATE $\theta\leftarrow d_k$
            \IF{$\theta$ is unvisited and $[\tilde{\mathbf{B}}]_{j\theta}^{t-1}>\min(\mathcal{B})$}
                \STATE $\mathcal{B}\leftarrow(\mathcal{B}\setminus\min(\mathcal{B}))\cup[\tilde{\mathbf{B}}]_{j\theta}^{t-1}$
            \ENDIF
            \STATE $\rho\leftarrow[\tilde{\mathbf{Q}}]_{i\kappa}+\nu_\theta^{t-1}$
            \STATE $\tilde{\mu}_j^t\leftarrow\sigma_{b(v_j)}(\mathcal{B})$
            \STATE $\tilde{\nu}_j^t\leftarrow\sigma_{b(v_j)+1}(\mathcal{B})$
            \STATE $k\leftarrow k+1$
        \ENDWHILE
        \STATE $\mu_j^t\leftarrow\tilde{\mu}_j^t$
        \STATE $\nu_j^t\leftarrow\tilde{\nu}_j^t$
    \end{algorithmic}
\end{algorithm}

\begin{rmk}
	Algorithm \ref{alg:Al1} is the key result of the proposed $b$-matching approach, which will be implemented in each Fog-AP. The procedure in each coherence time is listed as follows:
	\begin{enumerate}
		\item At the beginning of each coherence time, the user will feedback one bit CSI according to the channel estimation result to show if the Fog-AP is in outage for a user.
		\item Algorithm \ref{alg:Al1} will be executed between Fog-APs and users to select the Fog-APs  for each user. This algorithm will converge very quickly because of its low complexity.
		\item After Algorithm \ref{alg:Al1}, the data will be transmitted from the selected Fog-APs to each user.
  	\end{enumerate}
\end{rmk}

\section{Content Outage Performance Analysis and DMR Optimal MSR Code}\label{sec:outage_analysis}

In this section, the outage probability will be analyzed for problem P2. With the closed-form formula for content outage probability, the optimal $K_m$ can be obtained, so that the optimization problem P1 is solved.

\subsection{Conditional Content Outage Probability}

As discussed before, $\mathcal{M}_b^\mathrm{p}$ may not always exist, so that $k_m$ may be smaller than $K_m$ for user $u_m\in\mathcal{U}$. Thus, we need first calculate the tight upper bound of the content outage probability under the condition that $k_m$ non-outage Fog-APs can be accessed by user $u_m$. With a generated $\mathcal{M}_b^\mathrm{fm}$, the conditional content outage probability of user $u_m$ is written by
\begin{equation}\label{eq:con_outage_probability}
	p_m^\mathrm{con}(R_m|\mathcal{D}_{K_m k_m})=\Pr\left\{\left.\sum_{a_n\in\mathcal{A}_m}I_{mn}<R_m\right|\mathcal{D}_{K_m k_m}\right\},
\end{equation}
where
\begin{equation}\label{eq:state_space}
	\begin{aligned}
		\mathcal{D}_{K_m k_m}= & \{C_{m1},\ldots,C_{mK_m}|C_{m1}\geq\alpha_m^*,\ldots,C_{mk_m}\geq\alpha_m^*, \\
		& C_{m,k_m+1}<\alpha_m^*,\ldots,C_{mK_m}<\alpha_m^*\}.
	\end{aligned}
\end{equation}
In the following, the saddle-point approximation in \cite{butler_saddlepoint_2007,bai_conditional_2013} is applied to derive a tight upper bound of Eq. \eqref{eq:con_outage_probability}. Let $\rho_m=1-\frac{k_m}{K_m}$, and define a symbol ``$\lesssim$'' as
\begin{equation}\label{eq:symbol}
	f(x)\lesssim g(x)\Leftrightarrow
	\left\{
	\begin{aligned}
		& f(x)\leq g(x);\\
		& \lim_{x\to\infty}\frac{f(x)}{g(x)}=1.
	\end{aligned}
	\right.
\end{equation}
We then have the following theorem.

\medskip
\begin{lem}\label{lem:con_outage_exponent}
	The exponentially tight upper bound of $p_m^\mathrm{con}(R_m|\mathcal{D}_{K_m k_m})$ is given by
	\begin{equation}\label{eq:con_outage_exponent}
		\begin{aligned}
			& p_m^\mathrm{con}(R_m|\mathcal{D}_{K_m k_m})\lesssim p_m^\mathrm{upper}(R_m|\mathcal{D}_{K_m k_m}) \\
			& =\psi\exp\{-K_m[(\ln\gamma-\alpha_m^*)\lambda^*+J_1(\gamma)-J_0(\gamma)]\},
		\end{aligned}
	\end{equation}
	where
	\begin{equation}\label{eq:exponent_part1}
		\psi=\frac{1}{\sqrt{2\pi K_m\sigma^2}\lambda^*},
	\end{equation}
	\begin{equation}\label{eq:exponent_part2}
		\left\{
		\begin{aligned}
			J_1(\gamma)= & \ln p_m^{\rho_m}q_m^{1-\rho_m}; \\
			J_0(\gamma)= & \frac{1}{\gamma}+(1-\rho_m)\ln\Gamma\left(1-\lambda^*,\frac{e^{\alpha_m^*}}{\gamma}\right) \\
			& +\rho_m\ln\left(\Gamma\left(1-\lambda^*,\frac{1}{\gamma}\right)-\Gamma\left(1-\lambda^*,\frac{e^{\alpha_m^*}}{\gamma}\right)\right).
		\end{aligned}
		\right.
	\end{equation}
	In Eqs. \eqref{eq:exponent_part1} \eqref{eq:exponent_part2}, $\lambda^*$ and $\sigma^2$ satisfy Eq. \eqref{eq:exponent_equation} and Eq. \eqref{eq:exponent_parameter}, respectively. In these equations,
	\begin{equation}
		\Gamma(z,a)=\int_a^\infty e^{-t}t^{z-1}dt
	\end{equation}
	is the incomplete Gamma function, and
	\begin{equation}
		G_{p,q}^{m,n}\left(z\left|
		\begin{aligned}
			a_1, & \ldots,a_p \\
			b_1, & \ldots,b_q
		\end{aligned}
		\right.\right)=\frac{1}{2\pi i}\oint_\mathcal{L}\frac{\prod_{j=1}^m\Gamma(b_j-s)\prod_{j=1}^n\Gamma(1-a_j+s)}{\prod_{j=m+1}^q\Gamma(1-b_j+s)\prod_{j=n+1}^p\Gamma(a_j-s)}z^sds,
	\end{equation}
	is the Meijer's $G$-function \cite{gradshteyn_table_2007}.
\end{lem}
\begin{IEEEproof}
	See Appendix \ref{app:con_outage_exponent}.
\end{IEEEproof}

\begin{figure*}[h]
	{\scriptsize
	\begin{equation}\label{eq:exponent_equation}
		\begin{aligned}
			& \alpha_m^*= \ln\gamma+(1-\rho_m)\ln\frac{e^{\alpha_m^*}}{\gamma}+(1-\rho_m)\frac{G_{2,3}^{3,0}\left(\frac{e^{\alpha_m^*}}{\gamma}\left|
			\begin{array}{ccc}
				1, & 1, & - \\
				0, & 0, & 1-\lambda^*
			\end{array}\right.\right)}{\Gamma\left(1-\lambda^*,\frac{e^{\alpha_m^*}}{\gamma}\right)} \\
			 & +\rho_m\left[\frac{\Gamma\left(1-\lambda^*,\frac{1}{\gamma}\right)\ln\frac{1}{\gamma}-\Gamma\left(1-\lambda^*,\frac{e^{\alpha_m^*}}{\gamma}\right)\ln\frac{e^{\alpha_m^*}}{\gamma}+G_{2,3}^{3,0}\left(\frac{1}{\gamma}\left|
			 \begin{array}{ccc}
				1, & 1, & -\\
				0, & 0, & 1-\lambda^*
			\end{array}\right.\right)}{\Gamma\left(1-\lambda^*,\frac{1}{\gamma}\right)-\Gamma\left(1-\lambda^*,\frac{e^{\alpha_m^*}}{\gamma}\right)}\right. \\
			& \left.-\frac{G_{2,3}^{3,0}\left(\frac{e^{\alpha_m^*}}{\gamma}\left|
			\begin{array}{ccc}
				1, & 1, & - \\
				0, & 0, & 1-\lambda^*
			\end{array}\right.\right)}{\Gamma\left(1-\lambda^*,\frac{1}{\gamma}\right)-\Gamma\left(1-\lambda^*,\frac{e^{\alpha_m^*}}{\gamma}\right)}\right].
		\end{aligned}
	\end{equation}
	
	\begin{equation}\label{eq:exponent_parameter}
		\begin{aligned}
			& \sigma^2= \rho_m\left[\frac{\Gamma\left(1-\lambda^*,\frac{1}{\gamma}\right)\ln^2\frac{1}{\gamma}-\Gamma\left(1-\lambda^*,\frac{e^{\alpha_m^*}}{\gamma}\right)\ln^2\frac{e^{\alpha_m^*}}{\gamma}}{\Gamma\left(1-\lambda^*,\frac{1}{\gamma}\right)-\Gamma\left(1-\lambda^*,\frac{e^{\alpha_m^*}}{\gamma}\right)}\right. \\
			& +\frac{2G_{2,3}^{3,0}\left(\frac{1}{\gamma}\left|
			\begin{array}{ccc}
				1, & 1, & - \\
				0, & 0, & 1-\lambda^*
			\end{array}\right.\right)\ln\frac{1}{\gamma}-2G_{2,3}^{3,0}\left(\frac{e^{\alpha_m^*}}{\gamma}\left|
			\begin{array}{ccc}
				1, & 1, & - \\
				0, & 0, & 1-\lambda^*
			\end{array}\right.\right)\ln\frac{e^{\alpha_m^*}}{\gamma}}{\Gamma\left(1-\lambda^*,\frac{1}{\gamma}\right)-\Gamma\left(1-\lambda^*,\frac{e^{\alpha_m^*}}{\gamma}\right)} \\
			& \left.+\frac{2G_{3,4}^{4,0}\left(\frac{1}{\gamma}\left|
			\begin{array}{cccc}
				1, & 1, & 1, & - \\
				0, & 0, & 0, & 1-\lambda^*
			\end{array}\right.\right)-2G_{3,4}^{4,0}\left(\frac{e^{\alpha_m^*}}{\gamma}\left|
			\begin{array}{cccc}
				1, & 1, & 1, & - \\
				0, & 0, & 0, & 1-\lambda^*
			\end{array}\right.\right)}{\Gamma\left(1-\lambda^*,\frac{1}{\gamma}\right)-\Gamma\left(1-\lambda^*,\frac{e^{\alpha_m^*}}{\gamma}\right)}\right]\\ 
			& -\rho_m\left(\frac{\Gamma\left(1-\lambda^*,\frac{1}{\gamma}\right)\ln\frac{1}{\gamma}-\Gamma\left(1-\lambda^*,\frac{e^{\alpha_m^*}}{\gamma}\right)\ln\frac{e^{\alpha_m^*}}{\gamma}}{\Gamma\left(1-\lambda^*,\frac{1}{\gamma}\right)-\Gamma\left(1-\lambda^*,\frac{e^{\alpha_m^*}}{\gamma}\right)}\right. \\
			& +\left.\frac{G_{2,3}^{3,0}\left(\frac{1}{\gamma}\left|
			\begin{array}{ccc}
				1, & 1, & - \\
				0, & 0, & 1-\lambda^*
			\end{array}\right.\right)-G_{2,3}^{3,0}\left(\frac{e^{\alpha_m^*}}{\gamma}\left|
			\begin{array}{ccc}
				1, & 1, & - \\
				0, & 0, & 1-\lambda^*
			\end{array}\right.\right)}{\Gamma\left(1-\lambda^*,\frac{1}{\gamma}\right)-\Gamma\left(1-\lambda^*,\frac{e^{\alpha_m^*}}{\gamma}\right)}\right)^2+(1-\rho_m) \\
			& \cdot\left[\frac{2G_{3,4}^{4,0}\left(\frac{e^{\alpha_m^*}}{\gamma}\left|
			\begin{array}{cccc}
				1, & 1, & 1, & - \\
				0, & 0, & 0, & 1-\lambda^*
			\end{array}\right.\right)}{\Gamma\left(1-\lambda^*,\frac{e^{\alpha_m^*}}{\gamma}\right)}-\left(\frac{G_{2,3}^{3,0}\left(\frac{e^{\alpha_m^*}}{\gamma}\left|
			\begin{array}{ccc}
				1, & 1, & - \\
				0, & 0, & 1-\lambda^*
			\end{array}\right.\right)}{\Gamma\left(1-\lambda^*,\frac{e^{\alpha_m^*}}{\gamma}\right)}\right)^2\right].
		\end{aligned}
	\end{equation}
	}
\end{figure*}

\subsection{Content Outage Probability}

The content outage probability of the proposed RBG based $b$-matching approach is analyzed in this subsection. The exact performance analysis is intractable, so that we focus on the first order approximations in both high and low SNR regimes.

To analyze the content outage probability, we need to study the properties of the maximum $b$-matching on RBG first. Due to space limitation, we introduce the following lemma, where the detailed proof can be found in \cite{schrijver_combinatorial_2003},.

\medskip
\begin{lem}\label{lem:b_matching}
	\begin{equation}
		|\mathcal{M}_b^\mathrm{m}|=\min_{\mathcal{X}\subseteq\mathcal{V}}\{b(\mathcal{V}\setminus\mathcal{X})+|\mathcal{E}(\mathcal{X})|\}.
	\end{equation}
\end{lem}
\medskip

Let $\{K_{m:M}\}_{m=1}^M$ be the ordered sequence of $\{K_m\}_{m=1}^M$, where $K_{1:M}\leq\cdots\leq K_{M:M}$. Denote the ordered sequence of  $\{u_m\}_{m=1}^M$ as $\{u_{m:M}\}_{m=1}^M$, where $u_{m:M}$ has the same order as $K_{m:M}$. Define
\begin{equation}
	K^\mathrm{sum}=\sum_{m=1}^M K_m.
\end{equation}
The following lemma can then be established from Lemma \ref{lem:b_matching}.

\medskip
\begin{lem}\label{lem:matching_edges}
	Let $\mathcal{G}(\mathcal{U}\cup\mathcal{A},\mathcal{E})$ be a bipartite graph with $|\mathcal{U}|=M$ and $|\mathcal{A}|=N$. $b(v)$ is defined as Eq. \eqref{eq:b_v}. If $\mathcal{M}_b^\mathrm{m}$ exists such that:
	\begin{enumerate}
		\item the only unsaturated vertex is in $\{u_{i:M}\}_{i=m}^M$, the cardinality of $\mathcal{E}$ is given by
			\begin{equation}
				\Phi_1=(M-1)N+K_{m:M}-1;
			\end{equation}
		\item $\mathcal{U}$ has only one unsaturated vertex, the cardinality of $\mathcal{E}$ is given by
			\begin{equation}
				\Phi_2=(M-L)\left(\left\lceil\frac{K^\mathrm{sum}}{L}\right\rceil-1\right)+K^\mathrm{sum}-1.
			\end{equation}
	\end{enumerate}
\end{lem}
\begin{IEEEproof}
	See Appendix \ref{app:matching_edges}.
\end{IEEEproof}
\medskip

Based on Lemma \ref{lem:con_outage_exponent} and Lemma \ref{lem:matching_edges}, the content outage probability is summarized in the following theorem.

\medskip
\begin{thm}\label{thm:content_outage}
	If $N>\frac{\Phi_2-\eta K_m}{M-1}$, the content outage probability of $u_{m:M}$ in high SNRs is given by
	\begin{equation}\label{eq:outage_high1}
		p_{m:M}^\mathrm{out}(R_{m:M})=\sum_{\kappa=N-K_{m:M}+1}^N \binom{N}{\kappa}p_{m:M}^\mathrm{con}(R_{m:M}|\mathcal{D}_{K_{m:M},N-\kappa})p_{m:M}^\kappa+\mathsf{O}\left(p_{m:M}^N\right).
	\end{equation}
	If $N<\frac{\Phi_2-\eta K_m}{M-1}$, the content outage probability of $u_m$ in high SNRs is given by
	\begin{equation}\label{eq:outage_high2}
		p_m^\mathrm{out}(R_m) = \binom{MN}{\Phi_2}p_m^\mathrm{con}(R_m|\mathcal{D}_{K_m,\eta K_m})p_m^{MN-\Phi_2}+\mathsf{O}\left(p_m^{MN-\Phi_2+\eta K_m}\right).
	\end{equation}
	If $N=\frac{\Phi_2-\eta K_m}{M-1}$, the content outage probability of $u_{m:M}$ in high SNRs is given by the summation of Eq. \eqref{eq:outage_high1} and Eq. \eqref{eq:outage_high2}.
	
	In low SNRs, the content outage probability of $u_m$ is then given by
	\begin{equation}\label{eq:outage_low}
		p_m^\mathrm{out}(R_m) = p_m^N+Np_m^\mathrm{con}(R_m|\mathcal{D}_{K_m,1})p_m^{N-1}q_m+\mathsf{O}(q_m).
	\end{equation}
\end{thm}
\begin{IEEEproof}
	See Appendix \ref{app:content_outage}.
\end{IEEEproof}
\medskip

\subsection{DMR Optimal MSR Code}

\begin{figure}[t]
	\centering
	\begin{tikzpicture}
		\fill[domain=0:4,gray!50] plot (\x,{2*sqrt(1-pow(\x,2)/pow(4,2))}) -- (0,0) -- (0,2);
		\draw[domain=0:4,thick] plot (\x,{2*sqrt(1-pow(\x,2)/pow(4,2))});
		\node at (1.8,0.9) {$\mathcal{R}^\mathrm{OE}(R_1,R_2;\gamma)$};
		\node at (2.5,2.6) {OER Boundary};

		\draw[-latex] (2.5,2.4) -- (2,1.75);

		\draw[-latex,very thick] (0,0) -- (5,0) node[right] {$E_1(R_1)$};
		\draw[-latex,very thick] (0,0) -- (0,2.7) node[above] {$E_2(R_2)$};
	\end{tikzpicture}
	\caption{The OER $\mathcal{R}^\mathrm{OE}(R_1,R_2;\gamma)$ for a coded caching scheme in Fog-RAN.}\label{fig:region_example}
\end{figure}
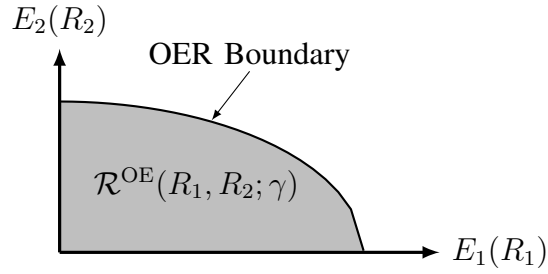

To illustrate the solution of P1 in Eq. \eqref{eq:MOPCO}, we define the \emph{outage exponent region} (OER), which specifies the set of content outage exponents that are simultaneously achievable by the users for the specific coded caching scheme. As a matter of fact, the outage exponent proposed in \cite{bai_outage_2013} is defined as
\begin{equation}\label{eq:outage_exponent}
	E_m(R_m,\gamma)=-\frac{\partial\ln p_m^\mathrm{out}(R_m)}{\partial\ln\gamma}.
\end{equation}
Similar to error exponent region in \cite{weng_error_2008}, the OER can be formally defined in the following.

\medskip
\begin{defn}\label{def:OER}
	For a coded caching scheme with $\mathbf{R}=(R_m)_{m=1}^M$, the OER, denoted by $\mathcal{R}^\mathrm{OE}(\mathbf{R};\gamma)$, consists of all vectors of outage exponents $(E_m(R_m,\gamma))_{m=1}^M$, which can be achieved by at least one feasible solution of Eq. \eqref{eq:b_matching_problem}.
\end{defn}
\medskip

\begin{figure}[t]
	\centering
	\includegraphics[width=3.6in]{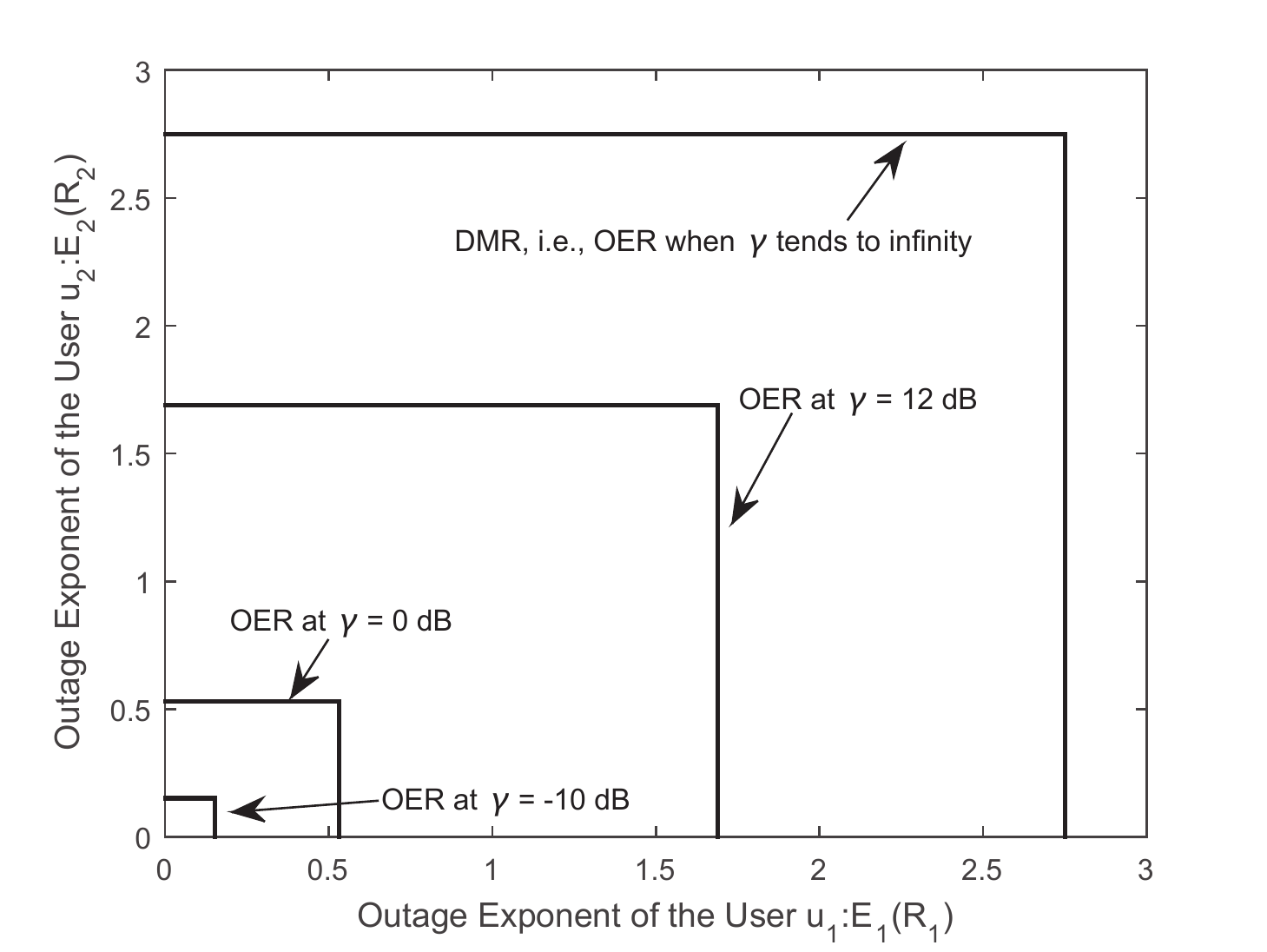}
	\caption{The OER and DMR for coded caching scheme in the Fog-RAN with $M=10$, $N=5$, $L=4$, and $\mathbf{r}_m=(0.9, 0.9)$.}\label{fig:OER}
\end{figure}

Fig. \ref{fig:region_example} shows an example of the OER for a coded caching scheme with $(R_1,R_2)$. In contrast to the unique capacity region, one OER corresponds to one coded caching scheme. Based on Theorem \ref{thm:content_outage}, the OER is summarized as follows. The OER at different SNRs are shown in Fig. \ref{fig:OER}.

\medskip
\begin{thm}\label{thm:OER}
	For a coded caching scheme with $\mathbf{R}$, the best achievable bound of the OER $\mathcal{R}_\mathrm{OE}(\mathbf{R};\gamma)$ is given by the vector $(E_m(R_m,\gamma))_{m=1}^M$, which can be obtained by plugging Eq. \eqref{eq:outage_high1} and Eq. \eqref{eq:outage_high2} into Eq. \eqref{eq:outage_exponent}.
\end{thm}
\medskip

The DMT for a user $u_m$ is closely related to the outage exponent as follows \cite{bai_outage_2013}:
\begin{equation}\label{eq:diversity_gain}
	\left\{
	\begin{aligned}
		& \lim_{\gamma\to\infty}E_m(R_m,\gamma)=d_m(r_m), \\
		& r_m=\lim_{\gamma\to\infty}\frac{R_m}{\ln\gamma},
	\end{aligned}
	\right.
\end{equation}
where $d_m(r_m)$ is the diversity gain for the multiplexing gain $r_m$. Recalling Eq. \eqref{eq:diversity_gain}, we have the following corollary.
\medskip
\begin{cor}\label{cor:CDMT}
	The conditional DMT for $u_m$ is given by
	\begin{equation}\label{eq:dmt}
		d_m^\mathrm{con}(r_m|\mathcal{D}_{K_mk_m})=K_m(1-\rho_m)\left(1-\frac{r_m}{K_m}\right).
	\end{equation}
\end{cor}
\medskip

Similar to OER, the DMT can be generalized to Fog-RANs, i.e., the \emph{diversity-multiplexing region} (DMR), which is defined in the following.

\medskip
\begin{defn}\label{def:DMR}
	For a coded caching scheme with $\mathbf{r}=(r_m)_{m=1}^M$, the DMR, denoted by $\mathcal{R}^\mathrm{DM}(\mathbf{r})$, consists of all vectors of diversity gains $(d_m(r_m))_{m=1}^M$, which can be achieved by at least one fesible solution of Eq. \eqref{eq:b_matching_problem}.
\end{defn}
\medskip

As shown in Fig. \ref{fig:region_example},  the DMR is the asymptotic case of the OER when $\gamma$ tends to infinity. According to Definition \ref{def:DMR} and Corollary \ref{cor:CDMT}, the achieved DMR is obtained from Theorem \ref{thm:content_outage}.

\medskip
\begin{thm}\label{thm:DMR}
	For a coded caching scheme with $\mathbf{r}=(r_m)_{m=1}^M$, the best achievable bound of the DMR $\mathcal{R}^\mathrm{DM}(\mathbf{r})$ is given by the vector $(d_m^*(r_{m}))_{m=1}^M$ for $0<r_m\leq K_m$ which satisfies
	\begin{equation}\label{eq:DMR}
		d_m^*(r_m)=N\left(1-\frac{r_m}{K_m}\right),
	\end{equation}
	if $N\geq\frac{\Phi_2-\eta K_m}{M-1}$; or
	\begin{equation}
		d_m^*(r_m)=(MN-\Phi_2+\eta K_m)\left(1-\frac{r_m}{K_m}\right),
	\end{equation}
	if $N<\frac{\Phi_2-\eta K_m}{M-1}$.
\end{thm}
\medskip

\begin{figure}[t]
	\centering
	\includegraphics[width=3.6in]{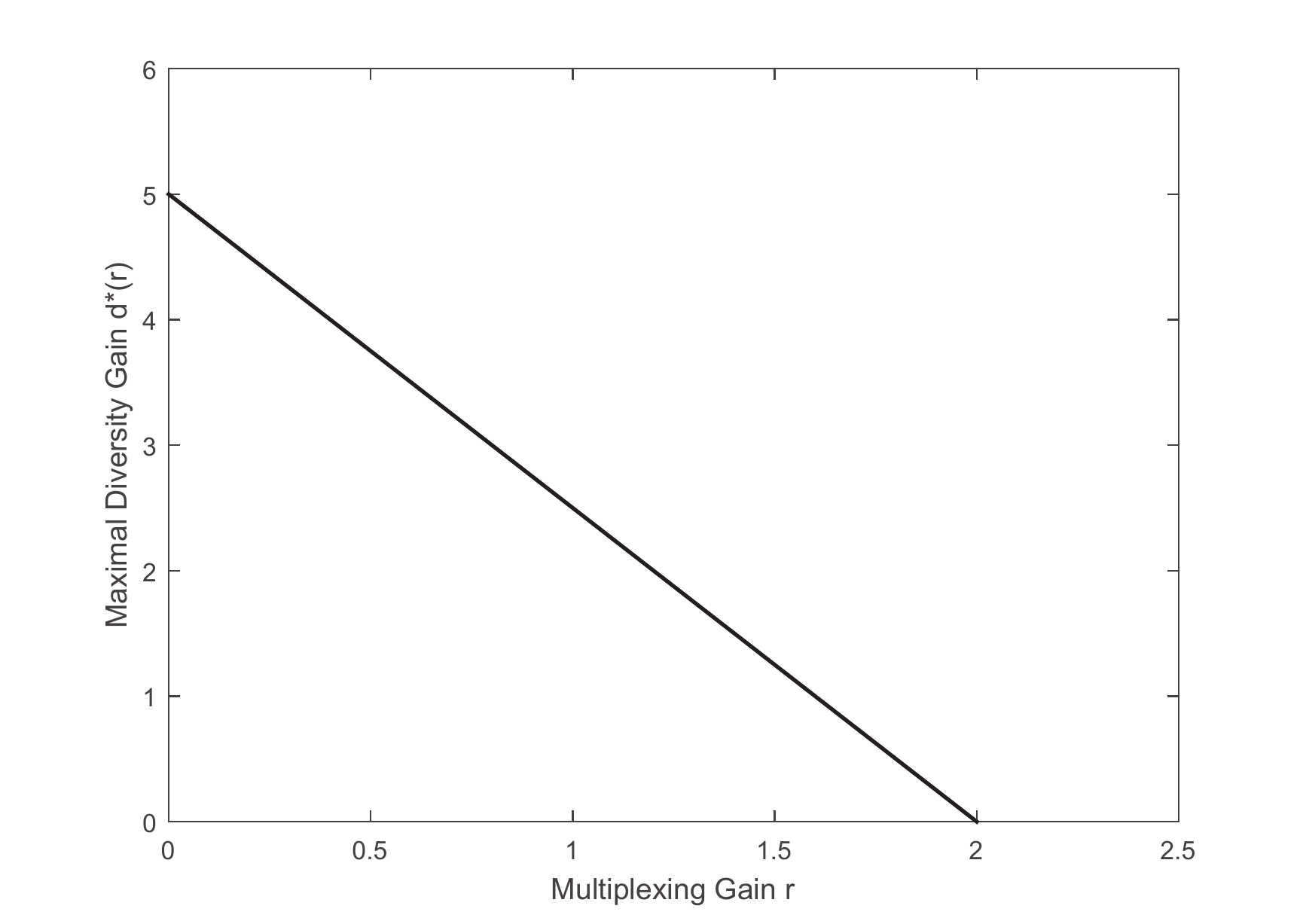}
	\caption{The DMT curve of user $u_m$ with $M=10$, $N=5$, $L=4$, and $r_m=0.9$.}\label{fig:DMTCurve}
\end{figure}

According to \cite{bai_outage_2013}, the optimal DMT for point-to-point parallel fading channels is given by
\begin{equation}
	d^*(r)=N\left(1-\frac{r}{N}\right),
\end{equation}
where $N$ is the number of subchannels. In this context, the optimal $K_m$, denoted by $K_m^*$, in P1 can be determined by
\begin{equation}\label{eq:OPTK_m}
	K_m^*=\frac{R_m}{\sum_{i=1}^M R_i}N.
\end{equation}
if the RBG based $b$-matching scheme is required to achieve the optimal DMR. Therefore, in the optimal DMR, each user shares the total multiplexing gain according to $\mathbf{R}$, while achieving the full frequency diversity. The DMR and the DMT curve of the user $u_m$ are respectively shown in Fig. \ref{fig:OER} and Fig. \ref{fig:DMTCurve}, where the OER approaches the DMR as SNR tends to infinity.

As discussed in Section \ref{subsec:CodedCaching}, the performance of MSR code is determined by $(\alpha_m^*,\beta_m^*)$. In this context, the requirement of the MSR code can be obtained if it needs to achieve the optimal DMR performance. According to \eqref{eq:OPTK_m}, the user $u_m$ will access $K_m^*=\frac{R_m}{\sum_{i=1}^M R_i}N$ Fog-APs to download the interested content $s_m$. Furthermore, recalling Eq. \eqref{eq:MSRcode}, we have
\begin{equation}
	\alpha_m^*=\frac{R_m}{K_m^*}=\frac{\sum_{i=1}^M R_i}{N}
\end{equation}
for $m=1,\ldots,M$. It can be seen that with the optimal $K_m^*$, $\alpha_m^*$, the information transmitted from a Fog-AP to user $u_m$, is the same as each other for the required content $s_m$ in $\mathcal{S}$. In Eq. \eqref{eq:MSRcode}, $\beta_m^*$ is given by
\begin{equation}
	\beta_m^*=\frac{D_m}{D_m-K_m^*+1}\frac{\sum_{i=1}^M R_i}{N}.
\end{equation}
Thus, the optimal $D_m$, denoted by $D_m^*$, is lower bounded by
\begin{equation}
	D_m^*\geq K_m^*=\frac{R_m}{\sum_{i=1}^M R_i}N.
\end{equation}
It can be seen that it requires at least $D_m^*=\frac{R_m}{\sum_{i=1}^M R_i}N$ Fog-APs to repair the content $s_m$ in a new Fog-AP. If we choose $D_m^*$ as the lower bound, the transmitted information is given by
\begin{equation}
	\beta_m^*=D_m^*\frac{\sum_{i=1}^M R_i}{N}=R_m.
\end{equation}
In this context, we have the following theorem.

\medskip
\begin{thm}\label{thm:MSRDesign}
	To achieve the optimal DMR performance of the coded caching scheme in Fog-RAN, the parameters of MSR code for $u_m$ with the content $s_m$ can be given by
	\begin{equation}
		(N,K_m^*,D_m^*)=\left(N,\frac{R_m}{\sum_{i=1}^M R_i}N,\frac{R_m}{\sum_{i=1}^M R_i}N\right).
	\end{equation}
	The code performance is characterized by
	\begin{equation}
		(\alpha_m^*,\beta_m^*)=\left(\frac{\sum_{i=1}^M R_i}{N},R_m\right).
	\end{equation}
\end{thm}
\medskip

The MSR code defined in Theorem \ref{thm:MSRDesign} achieves the Pareto optimal DMR with fairness constraints. Specifically, the fairness is guaranteed because the maximum achievable multiplexing gain for each user is proportional to the size of content stored in Fog-APs, i.e., $K_m\propto R_m$ for $m=1,\ldots,M$. According to Theorem \ref{thm:DMR}, for any given multiplexing gain $r_m\in(0,K_m]$, the $b$-matching approach achieves the maximum diversity gain. Moreover, any change of the MSR code parameters for all the users in $\mathcal{U}$ cannot improve the performance of one user without hurting another user, which fulfills the definition of Pareto optimal \cite{fudenberg_game_1991}.

\section{Simulation Results}\label{sec:simulation_results}

In this section, some simulation examples are presented. In the first group of simultions, the upper bound of the conditional outage probability is verified, which lays the foundation of the proposed framework. The second group of simulations verifies the outage performance. The last group of simulation compares the performance with different schemes and system parameters. In these figures, the DMT curves are plotted to illustrate that they are in parallel with the content outage probability curves (i.e., they have the same slope) in the high SNR regime.

\begin{figure}[t]
	\centering
	\includegraphics[width=3.6in]{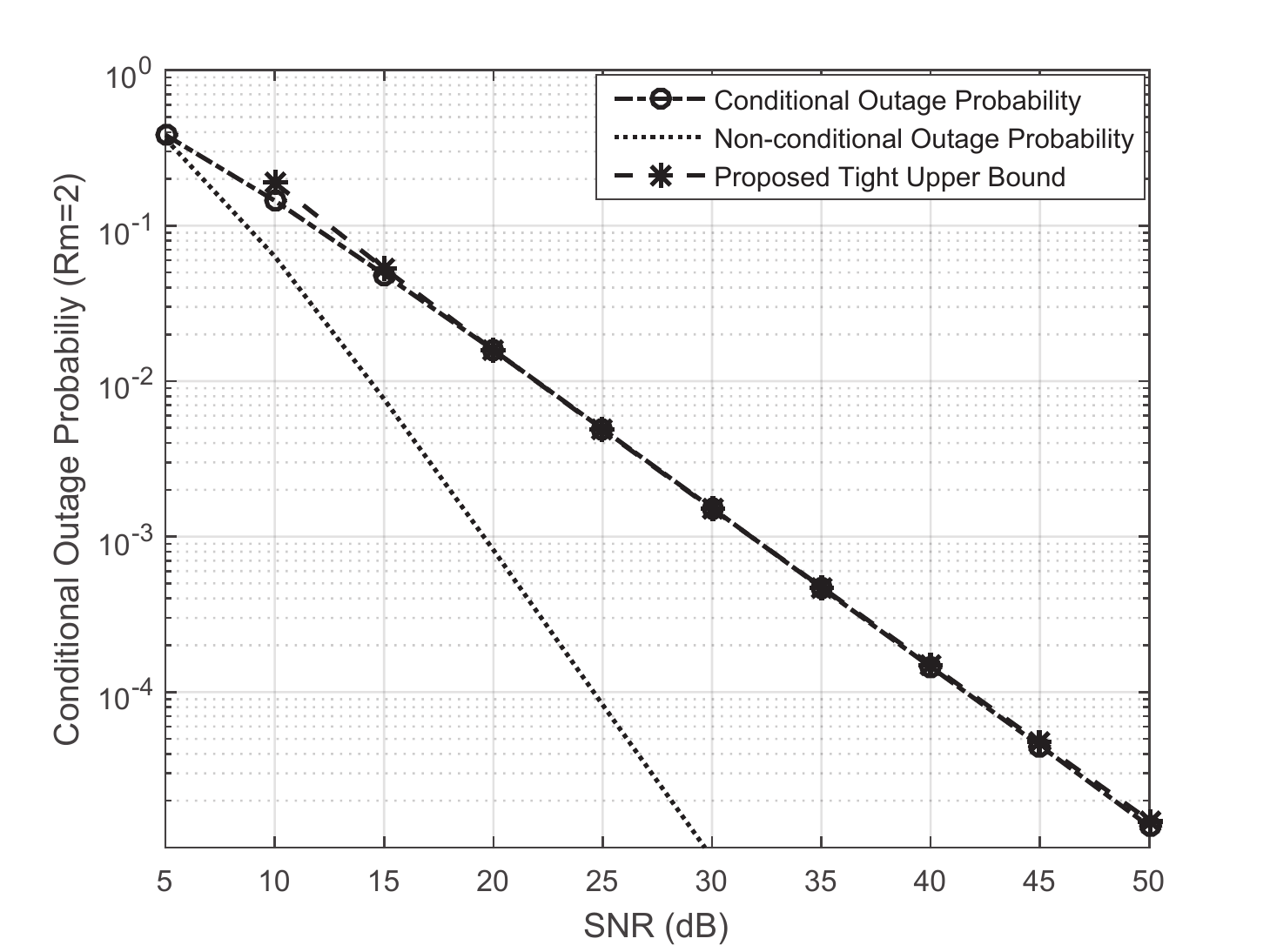}
	\caption{The tight upper bound of the conditional outage probability for the user $u_m$ with $K_m=2$, $k_m=1$, and $R_m=2$.}\label{fig:cond_outage_exponent}
\end{figure}

\begin{figure}[t]
	\centering
	\includegraphics[width=3.6in]{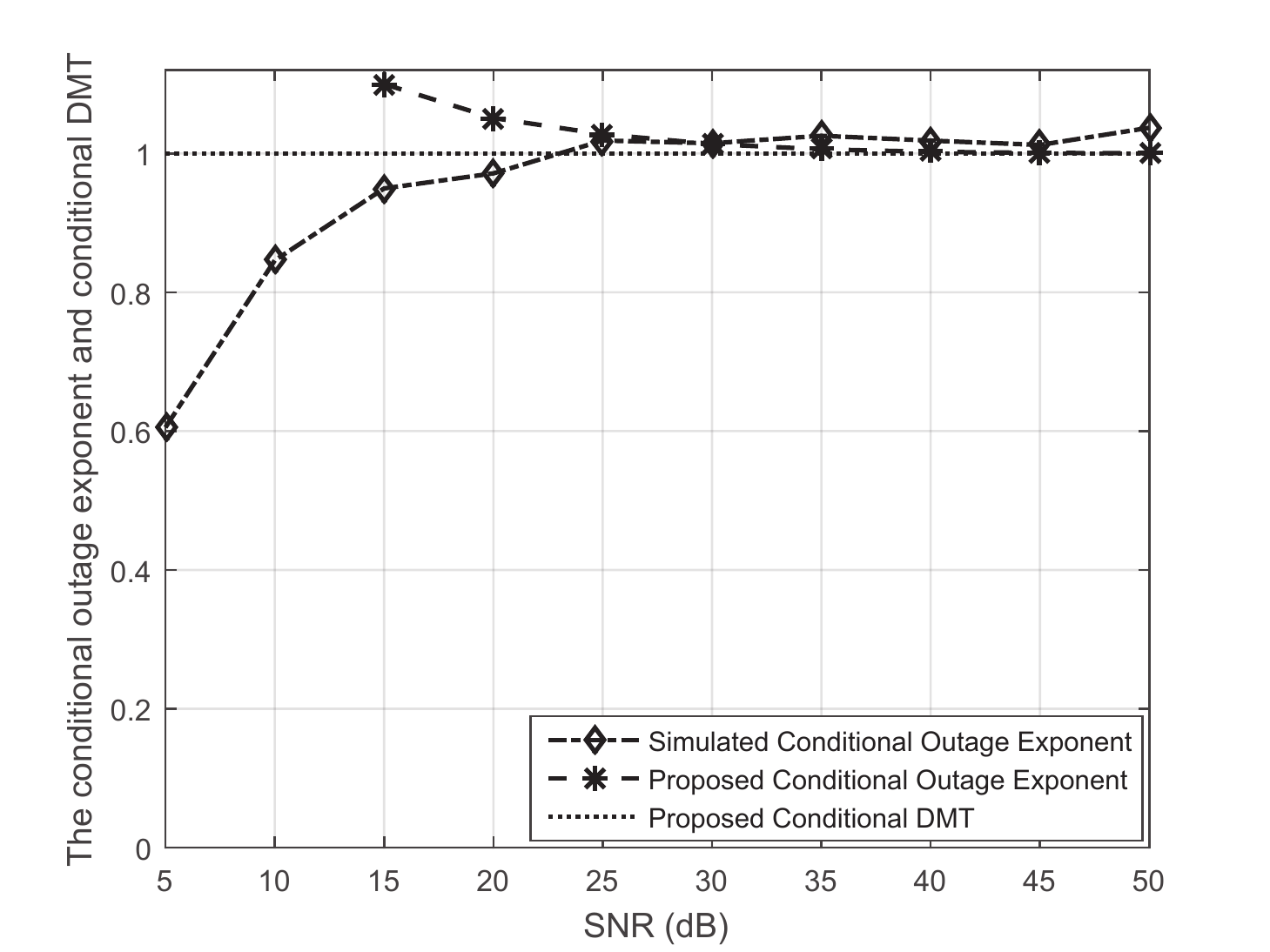}
	\caption{The conditional outage exponent and conditional DMT for the user $u_m$ with $K_m=2$, $k_m=1$, and $R_m=2$.}\label{fig:diversity_gain}
\end{figure}

For the first group of simulation results, a large number of samples are generated at each SNR value. The number of outage events can be obtained under the condition that the transmitter knows $\unit[1]{bit}$ CSI by computing the instantaneous channel capacity. The conditional outage probability $p^\mathrm{cop}_m\left(R_m\left|\mathcal{D}_{K_mk_m}\right.\right)$ is then calculated. The theoretical approximations are calculated by applying the results in Lemma \ref{lem:con_outage_exponent}. Fig. \ref{fig:cond_outage_exponent} compares the simulation results and the theoretical curves of the conditional outage probability. It can be seen that the proposed tight upper bound is nearly identical with the simulation results in high SNR regime. For comparison, the curve of non-conditional outage probability is also plotted. Fig. \ref{fig:diversity_gain} presents the curves of conditional outage exponent and conditional DMT at a given $R_m=2$. Clearly, the conditional outage exponent is an increasing function of SNR for a fixed $R_m$. The proposed theoretical curve is approaching the simulation one when SNR increases. In Fig. \ref{fig:diversity_gain}, the conditional DMT is plotted as a constant which is much larger than the outage exponent at low SNRs. However, the conditional outage exponent approaches the conditional DMT as SNR tends to infinity. Therefore, the proposed conditional outage exponent can be used to estimate the decreasing slope of conditional outage probabilities with $\unit[1]{bit}$ CSI.

\begin{figure}[t]
	\centering
	\includegraphics[width=3.6in]{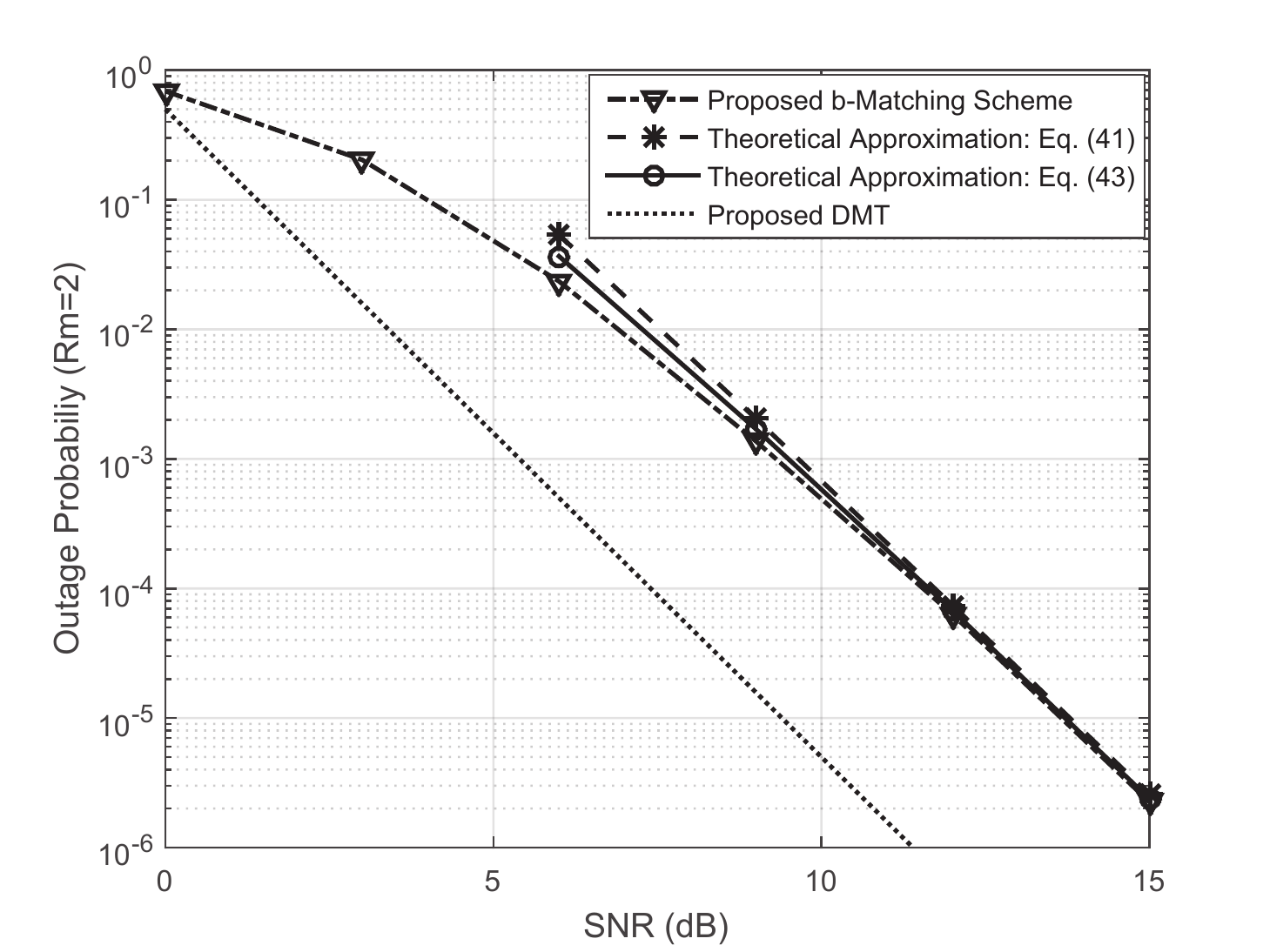}
	\caption{The outage probability of the $b$-matching approach for the user $u_m$ with $K_m=2$ and $R_m=2$ ($r_m=0$).}\label{fig:Rm2}
\end{figure}

\begin{figure}[t]
	\centering
	\includegraphics[width=3.6in]{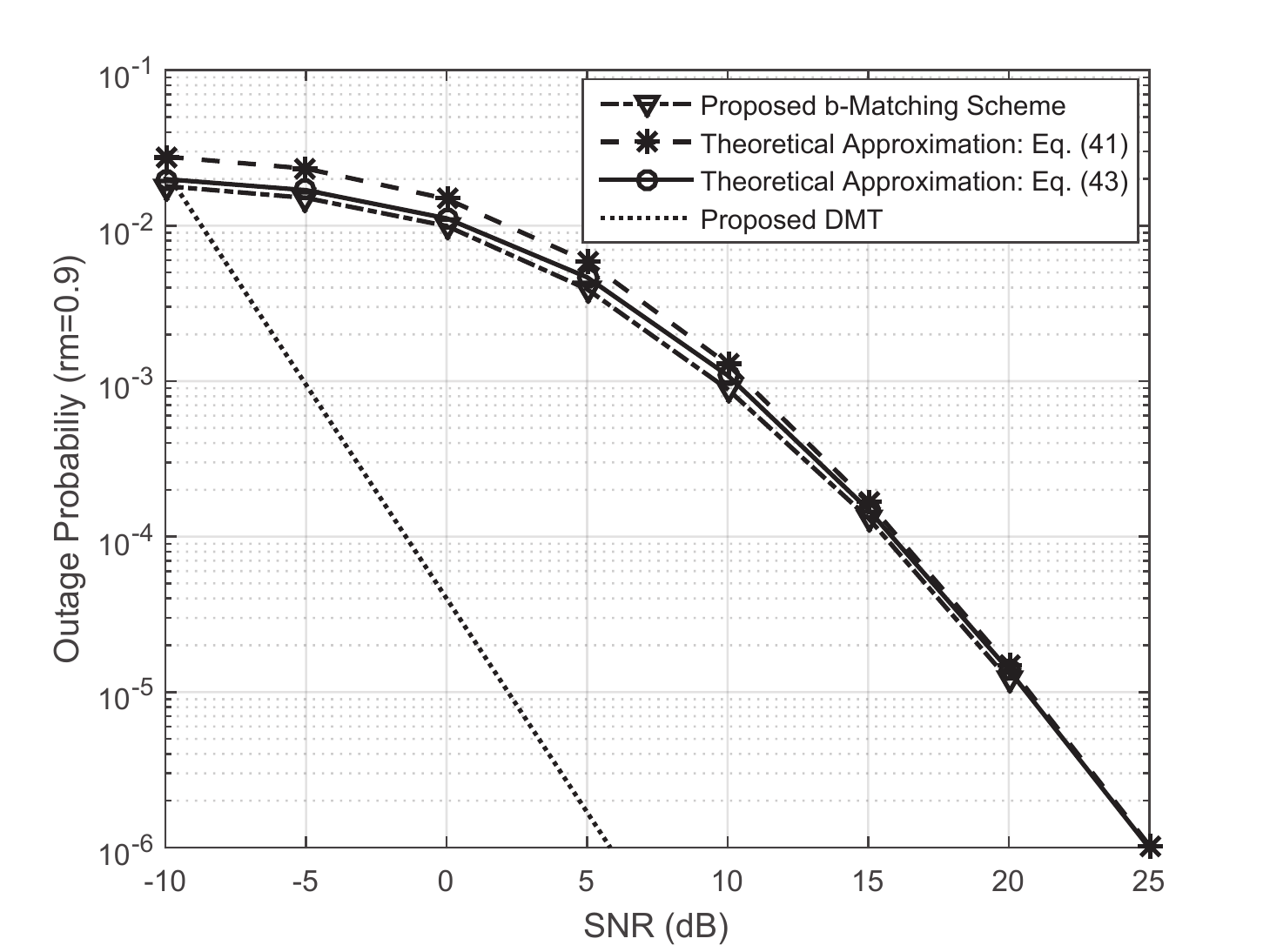}
	\caption{The outage probability of the $b$-matching approach for the user $u_m$ with $K_m=2$ and $r_m=0.9$.}\label{fig:rm0.9}
\end{figure}

\begin{figure}[t]
	\centering
	\includegraphics[width=3.6in]{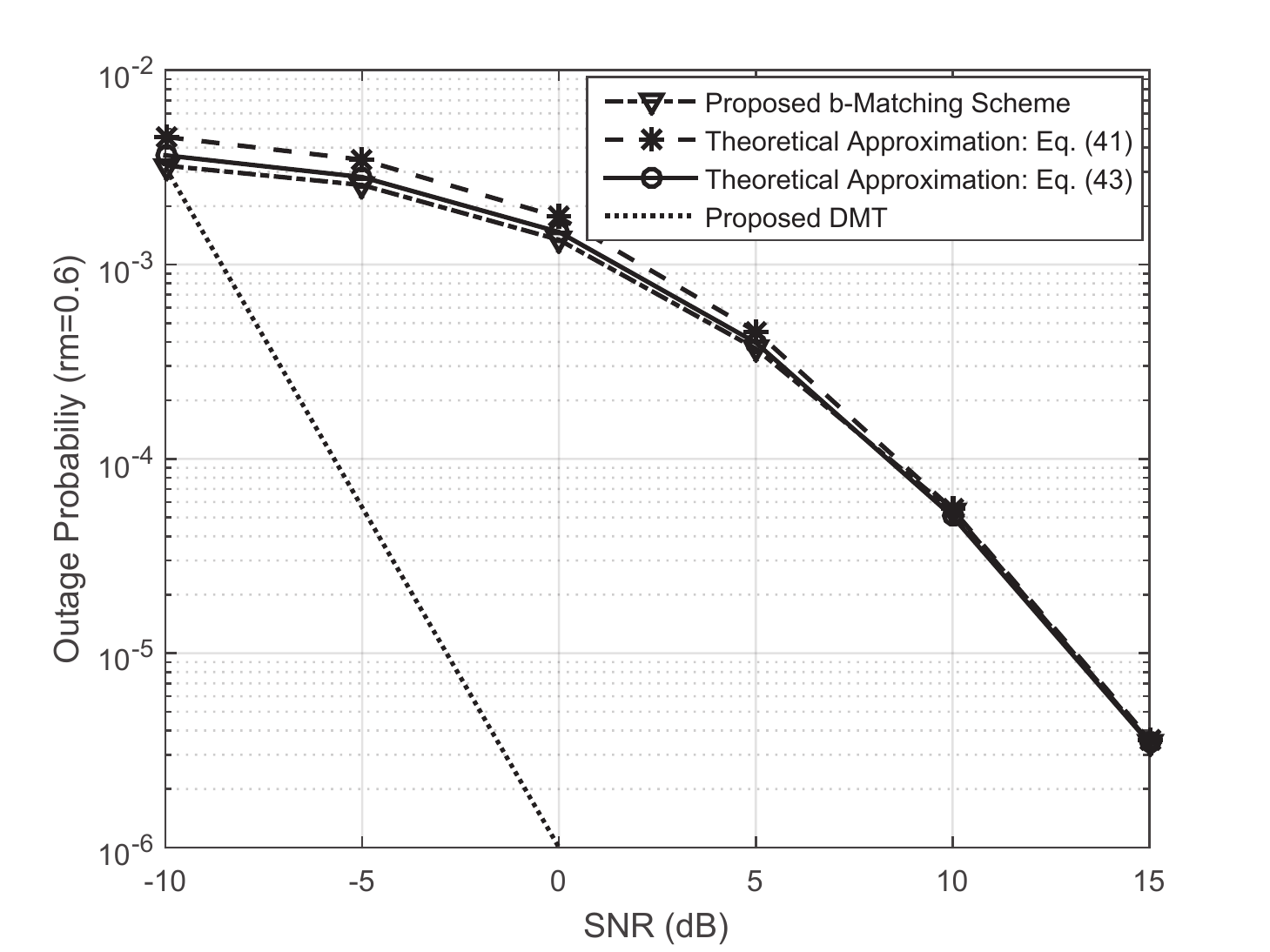}
	\caption{The outage probability of the $b$-matching approach for the user $u_m$ with $K_m=2$ and $r_m=0.6$.}\label{fig:rm0.6}
\end{figure}

For the second group, a large number of samples are generated according to the $b$-matching approach of the Fog-AP selection problem. In the proposed allocation scheme, the approximation of the outage probabilities are calculated by the formulas in Theorem \ref{thm:content_outage}. The asymptotic line $\mathsf{SNR}^{d_m^*(r_m)}$ is also plotted to compare the slope of these curves in high SNRs. Fig. \ref{fig:Rm2} compares the simulation results and the theoretical curves of the proposed $b$-matching approach. These simulation examples consider the Fog-AP selection problem with $M=10$ users and $N=5$ Fog-APs, where each Fog-AP can only support $L=4$ users at most and each user will access $K_m=2$ Fog-APs for fulfilling the requirement of MSR codes. In addition, the theoretical results of the content outage probability with Eq. \eqref{eq:outage_high1} and Eq. \eqref{eq:outage_low} are also illustrated simultaneously. The fixed target rate schemes, i.e., $R_m=2$, are plotted in Fig. \ref{fig:Rm2}. Since the conditional outage probability $p^\mathrm{cop}_m\left(R_m\left|\mathcal{D}_{K_mk_m}\right.\right)$ is calculated by utilizing the results in Lemma \ref{lem:con_outage_exponent}, the theoretical calculations are still suitable for only high SNR regime. It can be seen that the theoretical approximation in both Eq. \eqref{eq:outage_high1} and Eq. \eqref{eq:outage_low} approach the simulation results when SNR increases. Moreover, for a coded caching scheme with $\mathbf{r}=(r_m)_{m=1}^M=0$, the best achievable bound of the DMR $\mathcal{R}^\mathrm{DM}(\mathbf{r})$ given by the vector in Eq. \eqref{eq:DMR} satisfying $d_m^*(r_m)=5$ is also demonstrated in Fig. \ref{fig:Rm2}. In Fig. \ref{fig:rm0.9} and Fig. \ref{fig:rm0.6}, the dynamic rate scenario is considered, where the multiplexing gain $r_m$ is set to $0.9$ and $0.6$, respectively. The theoretical approximation is computed by Eq. \eqref{eq:outage_high1} and Eq. \eqref{eq:outage_low} in high and low SNR regimes, from $\unit[-10]{dB}$ to $\unit[25]{dB}$ under $r_m=0.9$ and from $\unit[-10]{dB}$ to $\unit[15]{dB}$ under $r_m=0.6$. The curves indicate that the theoretical approximation is nearly identical with the numeral simulation results, particularly for Eq. \eqref{eq:outage_low} in low SNR range. The best achievable bounds of the DMR $\mathcal{R}^\mathrm{DM}(\mathbf{r})$ calculated by Eq. \eqref{eq:DMR} are given as $d_m^*(r_m)=2.75$ and $d_m^*(r_m)=3.5$, respectively.

\begin{figure}[t]
	\centering
	\includegraphics[width=3.6in]{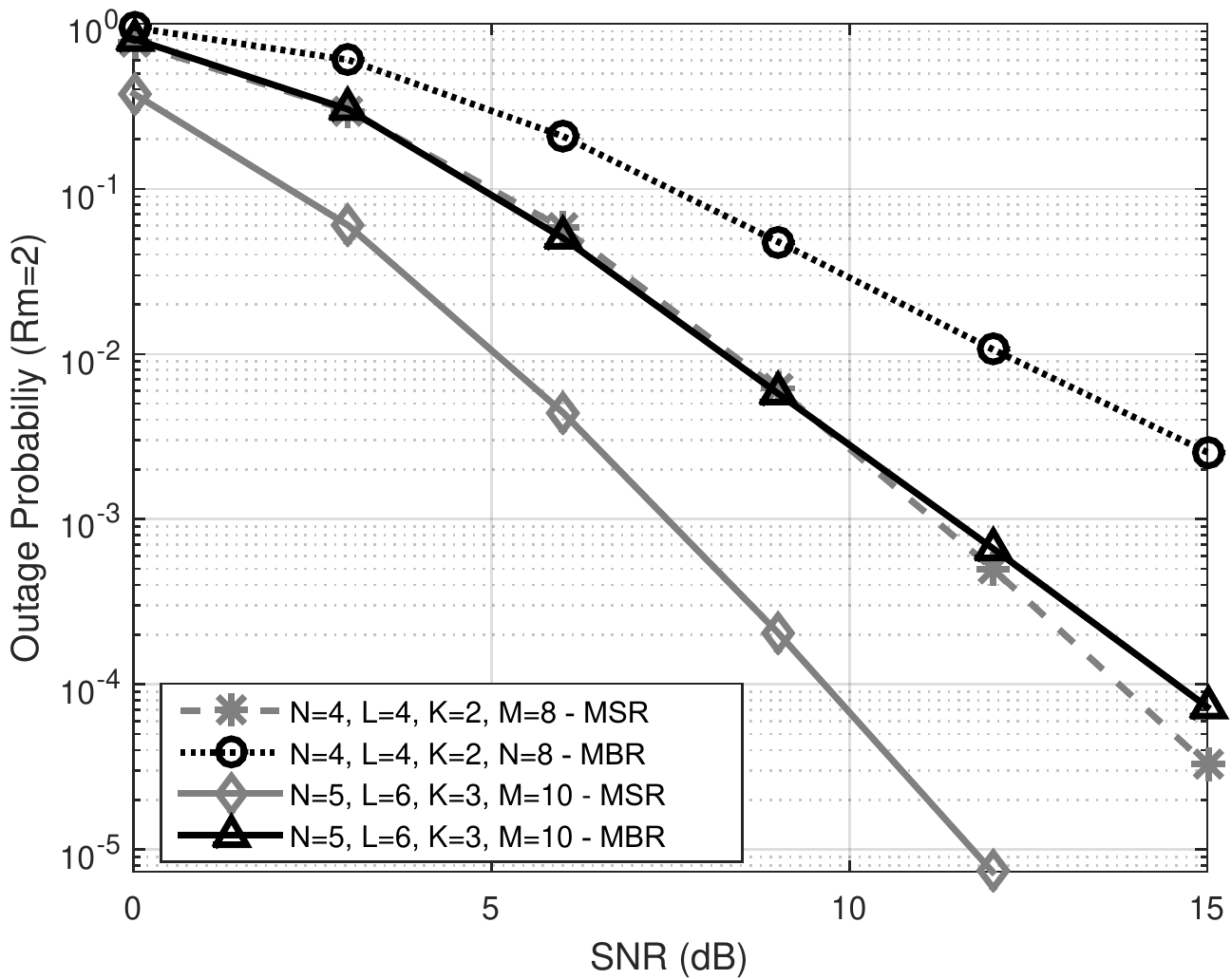}
	\caption{The outage probability of the $b$-matching approach for different coded caching schemes.}\label{fig:coding}
\end{figure}

\begin{figure}[t]
	\centering
	\includegraphics[width=3.6in]{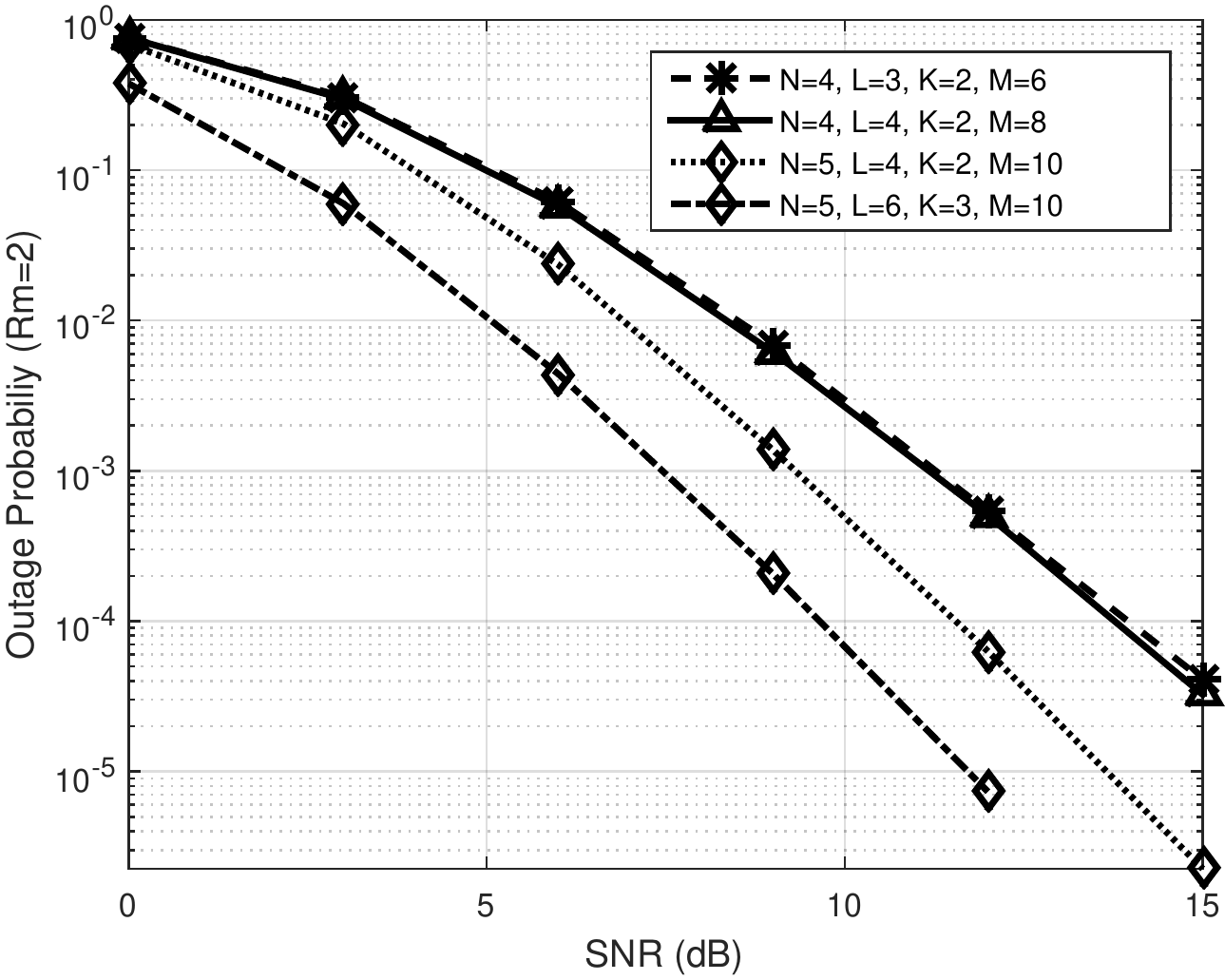}
	\caption{The outage probability of the $b$-matching approach for different system parameters.}\label{fig:parameters}
\end{figure}

The last group of simulation compares the performance with different schemes and system parameters. According to \cite{dimakis_survey_2011}, the commonly used coded caching schemes are summarized in Table \ref{tab:coding}. As the parameters of MDS code is quite different from our schemes. The results in Fig. \ref{fig:coding} show the performance comparison of MBR codes and MSR codes with different parameters. It can be seen that the outage probability of the MSR code outperforms the MBR code with the same coding parameters. Moreover, many parameters could influence the performance of the $b$-matching scheme. Hence, Fig. \ref{fig:parameters} shows the simulation results with different parameters. It can be seen that different parameters yield different outage performance. According to the theoretical analysis, the slope of the outage probability is determined in Theorem \ref{thm:DMR}. Therefore, the outage probability with $N=4,L=3,K_m=2,M=6$ is almost the same as that with $N=4,L=4,K_m=2,M=8$. However, the outage probability with $N=5,L=4,K_m=2,M=10$ is different form that with $N=5,L=6,K_m=3,M=10$ as $K_m$ is different in these curves.

\begin{table}[h]
	\caption{Parameters of Different Coding Schemes}\label{tab:coding}
	\centering
	\begin{tabular}{cccccc}
		\hline
		    & $n$ & $k$ & $d$     & $\alpha_m $                              & $\beta_m$ \\
		\hline
		MDS & $N$ & $K_m$ & $-$   & $\frac{R_m}{K_m}$                        & - \\
		MBR & $N$ & $K_m$ & $D_m$ & $\frac{2D_m}{2D_m-K_m+1}\frac{R_m}{K_m}$ & $\frac{2D_m}{2D_m-K_m+1} \frac{R_m}{K_m}$ \\
		MSR & $N$ & $K_m$ & $D_m$ & $\frac{R_m}{K_m}$                        & $\frac{D_m}{D_m-K_m+1}\frac{R_m}{K_m}$ \\
		\hline
	\end{tabular}
\end{table}

\section{Conclusions}\label{sec:conclusions}

This paper proposed a $b$-matching approach for coded cahing in Fog-RANs, which combines the advantages of coded caching scheme and RBG based fairness maximum $b$-matching. The combinatorial structure of Fog-AP selection is explored, and the optimal parameters of MSR code is obtained. The fast message passing algorithm is proposed to solve the optimal Fog-AP selection problem, which reduce the memory usesage to $O(M+N)$, compared to conventional message passing method. It was shown by theoretical derivations that all of the users fairly share the total multiplexing gain while achieving the full frequency diversity, i.e., the proposed scheme is optimal overall caching and transmission schemes from the DMR perspective. The simulation results illustrate the accuracy of the theoretical derivations, and verify that the optimal outage performance of the proposed scheme. Therefore, the $b$-matching approach provides an elegant theoretical framework for designing coded caching transmission scheme in Fog-RANs.

% if have a single appendix:
%\appendix[Proof of the Zonklar Equations]
% or
%\appendix  % for no appendix heading
% do not use \section anymore after \appendix, only \section*
% is possibly needed

% use appendices with more than one appendix
% then use \section to start each appendix
% you must declare a \section before using any
% \subsection or using \label (\appendices by itself
% starts a section numbered zero.)
%

\appendices

\section{Preliminaries on RBG and $b$-Matching}\label{app:RBG_b_matching}

In this appendix, we introduce some fundamental concepts of RBG \cite{bollobas_random_2001}. A bipartite graph $\mathcal{G}(\mathcal{V}_1\cup\mathcal{V}_2,\mathcal{E})$ is a graph on which the set of vertices $\mathcal{V}$ admits a partition into two classes $\mathcal{V}_1$ and $\mathcal{V}_2$ such that every edge in the edge set $\mathcal{E}$ has its ends in different classes. Specifically, for any edge $e=(v,v')$ in $\mathcal{E}$, we have $v\in\mathcal{V}_1$ and $v'\in\mathcal{V}_2$. For convenience, denote $v\in e$ if $e=(v,v')$. A bipartite graph with $M$ and $N$ vertices in different partition classes is called complete, denoted by $\mathcal{K}_{MN}$, if every two vertices from different partition classes are adjacent. Similar to \cite{bollobas_random_2001}, the probability space of a RBG is defined by $\mathscr{G}\{\mathcal{K}_{MN};\mathsf{P}\}$, which means that whether the edge of $\mathcal{K}_{MN}$ appears follows the probability measure $\mathsf{P}$. A vertex $v'$ is called a neighbor of vertex $v$, if they are adjacent. All of the neighboring vertices of a given vertex $v$ are denoted by the set $\mathcal{N}(v)\subset\mathcal{V}$. The cardinality of $\mathcal{N}(v)$ is referred to as the degree of the vertex $v$, denoted by $\deg_\mathcal{G}(v)$. The adjacent vertex set of $\mathcal{X}\subseteq\mathcal{V}$ is then defined by
\begin{equation}
	\mathcal{N}(\mathcal{X})=\bigcup_{v\in\mathcal{X}}\mathcal{N}(v).
\end{equation}
For convenience, we also denote
\begin{equation}
	\mathcal{E}(\mathcal{X})=\bigcup_{v\in\mathcal{X}}\mathcal{E}(v),
\end{equation}
and
\begin{equation}
	b(\mathcal{X})=\bigcup_{v\in\mathcal{X}}b(v)
\end{equation}
for any $\mathcal{X}\subseteq\mathcal{V}$.

Based on the bipartite graph, the $b$-matching can be defined as follows \cite{schrijver_combinatorial_2003}. Let $b(v)$ be a non-negative integer associated with the vertex $v\in\mathcal{V}$, and $\mathcal{E}(v)$ be the set of edges incident with $v$. The $b$-matching $\mathcal{M}_b$ is the subset of $\mathcal{E}$ satisfying
\begin{equation}\label{eq:b_matching}
	|\mathcal{M}_b\cap\mathcal{E}(v)|\leq b(v).
\end{equation}
If $|\mathcal{M}_b\cap\mathcal{E}(v)|=b(v)$, the vertex $v$ is referred to as the $\mathcal{M}_b$-saturated. The maximum $b$-matching, denoted by $\mathcal{M}_b^\mathrm{m}$, is the $b$-matching with the maximum number of edges. Especially, the $\mathcal{V}_1$-perfect $b$-matching, denoted by $\mathcal{M}_b^\mathrm{p}$, means that the equality of Eq. \eqref{eq:b_matching} holds for every $v\in\mathcal{V}_1$.

\section{Proof of Lemma \ref{lem:con_outage_exponent}}\label{app:con_outage_exponent}

Define a sequence of random variables $\{Z_{mn}\}_{n=1}^{K_m}$. The cumulative distribution function of $Z_{mn}$ for $1\leq n\leq k_m$ is given by
\begin{equation}
	\begin{aligned}
		F_{Z_{mn}}(z) & =\Pr\{I_{mn}<z|\mathcal{D}_{K_m k_m}\}=\Pr\{I_{mn}<z|I_{mn}\geq\alpha_m^*\} \\ 
		& =\left\{
		\begin{aligned}
			& 1-\frac{1}{q_m}\exp\left(-\frac{e^z-1}{\gamma}\right),	 && z\geq\alpha_m^*; \\
			& 0, && \text{otherwise}.
		\end{aligned}
		\right.
	\end{aligned}
\end{equation}
For $k_m+1\leq n\leq K_m$, the cumulative distribution function of $Z_{mn}$ is given by
\begin{equation}
	\begin{aligned}
		F_{Z_{mn}}(z) & =\Pr\{I_{mn}<z|\mathcal{D}_{K_m k_m}\}=\Pr\{I_{mn}<z|I_{mn}<\alpha_m^*\} \\
		& =\left\{
		\begin{aligned}
			& \frac{1}{p_m}\left[1-\exp\left(-\frac{e^z-1}{\gamma}\right)\right],	 && z<\alpha_m^*; \\
			& 1, && \text{otherwise}.
		\end{aligned}
		\right.
	\end{aligned}
\end{equation}
Let $X_n=\alpha_m^*-Z_{mn}$ and $Y_{K_m}=\sum_{n=1}^{K_m}X_n$, the conditional outage probability in Eq. \eqref{eq:con_outage_probability} can be rewritten as $p_m^\mathrm{con}(R_m|\mathcal{D}_{K_m k_m})=\Pr\{Y_{K_m}>0\}$.
%\begin{equation}
%	p_m^\mathrm{con}(R_m|\mathcal{D}_{K_m k_m})=\Pr\{Y_{K_m}>0\}.
%\end{equation}

Recalling that the elements of $\{h_{mn}\}_{n=1}^{K_m}$ are independent with the identical distribution of $\mathcal{CN}(0,1)$. Let $1\leq n_1\leq k_m$, $k_m+1\leq n_2\leq K_m$, the cumulant-generating function of $Y_{K_m}$ is then given by
\begin{equation}
%	\begin{aligned}
%		& \Lambda(\lambda)=\lim_{K_m\to\infty}\frac{1}{K_m}\ln\mathbb{E}\left\{e^{\lambda Y_{K_m}}\right\} \\
%		= & \alpha_m^*\lambda+\ln\left(\mathbb{E}\left\{e^{-\lambda Z_{mn_2}}\right\}\right)^{\rho_m}+\ln\left(\mathbb{E}\left\{e^{-\lambda Z_{mn_1}}\right\}\right)^{1-\rho_m} \\
%		= & (\alpha_m^*-\ln\gamma)\lambda+\frac{1}{\gamma}-\ln p_m^{\rho_m}q_m^{1-\rho_m}+(1-\rho_m)\ln\Gamma\left(1-\lambda,\frac{e^{\alpha_m^*}}{\gamma}\right)+ \\
%		& \rho_m\ln\left(\Gamma\left(1-\lambda,\frac{1}{\gamma}\right)-\Gamma\left(1-\lambda,\frac{e^{\alpha_m^*}}{\gamma}\right)\right)
%	\end{aligned}
	\begin{aligned}
		& \Lambda(\lambda)=\lim_{K_m\to\infty}\frac{1}{K_m}\ln\mathbb{E}\left\{e^{\lambda Y_{K_m}}\right\} = \alpha_m^*\lambda+\ln\left(\mathbb{E}\left\{e^{-\lambda Z_{mn_2}}\right\}\right)^{\rho_m}+\ln\left(\mathbb{E}\left\{e^{-\lambda Z_{mn_1}}\right\}\right)^{1-\rho_m} \\
		= & (\alpha_m^*-\ln\gamma)\lambda+\frac{1}{\gamma}-\ln p_m^{\rho_m}q_m^{1-\rho_m}+(1-\rho_m)\ln\Gamma\left(1-\lambda,\frac{e^{\alpha_m^*}}{\gamma}\right)+ \\
		& \rho_m\ln\left(\Gamma\left(1-\lambda,\frac{1}{\gamma}\right)-\Gamma\left(1-\lambda,\frac{e^{\alpha_m^*}}{\gamma}\right)\right)
	\end{aligned}
\end{equation}
where $\rho_m=1-\frac{k_m}{K_m}$. Considering the relation between the cumulant-generating function and the characteristic function, the characteristic function of $Y_{K_m}$ can then be given by $\Phi_{K_m}(i\lambda)=K_m\Lambda(i\lambda)$.

According to L\'evy's theorem, we have
\begin{equation}
%	\begin{aligned}
%		G(y)= & \Pr\{Y_{K_m}>y\} \\
%		= & \frac{1}{2\pi i}\int_{-\infty}^{+\infty}\frac{1}{\zeta}\exp(K_m\Lambda(i\zeta)-i\zeta y)d\zeta \\
%		= & \frac{1}{2\pi i}\int_{\lambda-i\infty}^{\lambda+i\infty}\frac{1}{z}\exp(K_m\Lambda(z)-zy)dz,
%	\end{aligned}
	\begin{aligned}
		G(y)= & \Pr\{Y_{K_m}>y\} = \frac{1}{2\pi i}\int_{-\infty}^{+\infty}\frac{1}{\zeta}\exp(K_m\Lambda(i\zeta)-i\zeta y)d\zeta \\
		= & \frac{1}{2\pi i}\int_{\lambda-i\infty}^{\lambda+i\infty}\frac{1}{z}\exp(K_m\Lambda(z)-zy)dz,
	\end{aligned}
\end{equation}
where $z=\lambda+i\zeta$, and $\lambda$ is chosen from the convergence region of this integral. According to the saddle-point approximation method \cite{butler_saddlepoint_2007}, if we let $z^*$ with $\Re(z^*)=\lambda^*$ be a solution of the saddle-point equation $
	\Lambda'(z)=\frac{y}{K_m}$, then
\begin{equation}
	\frac{1}{2\pi i}\int_{\lambda^*-i\infty}^{\lambda^*+i\infty}\frac{e^{K_m\Lambda(z)-zy}}{z}dz\lesssim\frac{e^{K_m\Lambda(z^*)-z^*y}}{\sqrt{2\pi K_m\Lambda''(\lambda^*)}\lambda^*}.
\end{equation}
Since $G(y)$ is a real function, we can always choose a real $z^*$ only if the cumulant-generating function exists. Moreover, $\frac{\lambda y}{K_m}-\Lambda(\lambda)$ is a conjugate function when $\lambda\in\mathbb{R}$, then $\lambda^*$ must be unique. Therefore, the exponentially tight upper bound of $p_m^\mathrm{con}(R_m|\mathcal{D}_{K_m k_m})$ is the tail distribution of $G(y)$ with $y>\mathbb{E}\{Y_{K_m}\}$, which is given by
\begin{equation}\label{eq:saddle-point}
%	\begin{aligned}
%		& p_m^\mathrm{con}(R_m|\mathcal{D}_{K_m k_m})=G(0) \\ 
%		& \lesssim\frac{1}{\sqrt{2\pi K_m\sigma^2}\lambda^*}e^{K_m\Lambda(\lambda^*)}=p_m^{\mathrm{upper}}(R_m|\mathcal{D}_{K_m k_m}),
%	\end{aligned}
	p_m^\mathrm{con}(R_m|\mathcal{D}_{K_m k_m})=G(0)\lesssim\frac{1}{\sqrt{2\pi K_m\sigma^2}\lambda^*}e^{K_m\Lambda(\lambda^*)}=p_m^{\mathrm{upper}}(R_m|\mathcal{D}_{K_m k_m}),
\end{equation}
where $\lambda^*$ is the solution of $\Lambda'(\lambda)=0$, and $\sigma^2=\Lambda''(\lambda^*)$. Clearly, we have
\begin{equation}\label{eq:CGF_derivative}
	\begin{aligned}
		\Lambda'(\lambda)= & (\alpha_m^*-\ln\gamma)+\left(1-\rho_m\right)\frac{\Gamma'\left(1-\lambda,\frac{e^{\alpha_m^*}}{\gamma}\right)}{\Gamma\left(1-\lambda,\frac{e^{\alpha_m^*}}{\gamma}\right)} \\
		& +\rho_m\frac{\Gamma'\left(1-\lambda,\frac{1}{\gamma}\right)-\Gamma'\left(1-\lambda,\frac{e^{\alpha_m^*}}{\gamma}\right)}{\Gamma\left(1-\lambda,\frac{1}{\gamma}\right)-\Gamma\left(1-\lambda,\frac{e^{\alpha_m^*}}{\gamma}\right)}=0.
	\end{aligned}
\end{equation}
According to the properties of Gamma function and Meijer's $G$-function, we have
\begin{equation}\label{eq:gamma_derivative}
%	\begin{aligned}
%		\Gamma'(1-\lambda,z)= & -\Gamma(1-\lambda,z)\ln z \\
%		& -G_{2,3}^{3,0}\left(z\left|
%		\begin{array}{ccc}
%			1, & 1, & - \\
%			0, & 0, & 1-\lambda
%		\end{array}
%		\right.\right).
%	\end{aligned}
	\Gamma'(1-\lambda,z) = -\Gamma(1-\lambda,z)\ln z
	-G_{2,3}^{3,0}\left(z\left|
	\begin{array}{ccc}
		1, & 1, & - \\
		0, & 0, & 1-\lambda
	\end{array}
	\right.\right).
\end{equation}
Therefore, Eq. \eqref{eq:exponent_equation} can be obtained by plugging Eq. \eqref{eq:gamma_derivative} into Eq. \eqref{eq:CGF_derivative}. For $\sigma^2$, we have
\begin{equation}
%	\begin{aligned}\label{eq:CGF_parameter}
%		& \sigma^2=\Lambda''(\lambda^*) \\
%		& = (1-\rho_m)\left[\frac{\Gamma''\left(1-\lambda^*,\frac{e^{\alpha_m^*}}{\gamma}\right)}{\Gamma\left(1-\lambda^*,\frac{e^{\alpha_m^*}}{\gamma}\right)}-\left(\frac{\Gamma'\left(1-\lambda^*,\frac{e^{\alpha_m^*}} {\gamma}\right)}{\Gamma\left(1-\lambda^*,\frac{e^{\alpha_m^*}}{\gamma}\right)}\right)^2\right] \\
%		& +\rho_m\frac{\Gamma''\left(1-\lambda^*,\frac{1}{\gamma}\right)-\Gamma''\left(1-\lambda^*,\frac{e^{\alpha_m^*}}{\gamma}\right)}{\Gamma\left(1-\lambda^*,\frac{1}{\gamma}\right)-\Gamma\left(1-\lambda^*,\frac{e^{\alpha_m^*}}{\gamma}\right)} \\
%		& -\rho_m\left(\frac{\Gamma'\left(1-\lambda^*,\frac{1}{\gamma}\right)-\Gamma'\left(1-\lambda^*,\frac{e^{\alpha_m^*}}{\gamma}\right)}{\Gamma\left(1-\lambda^*,\frac{1}{\gamma}\right)-\Gamma\left(1-\lambda^*,\frac{e^{\alpha_m^*}}{\gamma}\right)}\right)^2.
%	\end{aligned}
\begin{aligned}\label{eq:CGF_parameter}
	& \sigma^2=\Lambda''(\lambda^*) = (1-\rho_m)\left[\frac{\Gamma''\left(1-\lambda^*,\frac{e^{\alpha_m^*}}{\gamma}\right)}{\Gamma\left(1-\lambda^*,\frac{e^{\alpha_m^*}}{\gamma}\right)}-\left(\frac{\Gamma'\left(1-\lambda^*,\frac{e^{\alpha_m^*}} {\gamma}\right)}{\Gamma\left(1-\lambda^*,\frac{e^{\alpha_m^*}}{\gamma}\right)}\right)^2\right] \\
	& +\rho_m\frac{\Gamma''\left(1-\lambda^*,\frac{1}{\gamma}\right)-\Gamma''\left(1-\lambda^*,\frac{e^{\alpha_m^*}}{\gamma}\right)}{\Gamma\left(1-\lambda^*,\frac{1}{\gamma}\right)-\Gamma\left(1-\lambda^*,\frac{e^{\alpha_m^*}}{\gamma}\right)}-\rho_m\left(\frac{\Gamma'\left(1-\lambda^*,\frac{1}{\gamma}\right)-\Gamma'\left(1-\lambda^*,\frac{e^{\alpha_m^*}}{\gamma}\right)}{\Gamma\left(1-\lambda^*,\frac{1}{\gamma}\right)-\Gamma\left(1-\lambda^*,\frac{e^{\alpha_m^*}}{\gamma}\right)}\right)^2.
\end{aligned}
\end{equation}
According to the properties of Meijer's $G$-function, we have
\begin{equation}\label{eq:Meijer_derivative}
%	\begin{aligned}
%		& \frac{\partial}{\partial x}G_{2,3}^{3,0}\left(z\left|
%		\begin{array}{ccc}
%			1, & 1, & - \\
%			0, & 0, & 1-\lambda^*
%		\end{array}
%		\right.\right) \\
%		= & G_{2,3}^{3,0}\left(z\left|
%		\begin{array}{ccc}
%			1, & 1, & - \\
%			0, & 0, & 1-\lambda^*
%		\end{array}\right.\right)\ln z \\
%		& +2G_{3,4}^{4,0}\left(z\left|
%		\begin{array}{cccc}
%			1, & 1, & 1, & - \\
%			0, & 0, & 0, & 1-\lambda^*
%		\end{array}
%		\right.\right)
%	\end{aligned}
	\begin{aligned}
		& \frac{\partial}{\partial x}G_{2,3}^{3,0}\left(z\left|
		\begin{array}{ccc}
			1, & 1, & - \\
			0, & 0, & 1-\lambda^*
		\end{array}
		\right.\right) \\
		= & G_{2,3}^{3,0}\left(z\left|
		\begin{array}{ccc}
			1, & 1, & - \\
			0, & 0, & 1-\lambda^*
		\end{array}\right.\right)\ln z
		+2G_{3,4}^{4,0}\left(z\left|
		\begin{array}{cccc}
			1, & 1, & 1, & - \\
			0, & 0, & 0, & 1-\lambda^*
		\end{array}
		\right.\right)
	\end{aligned}
\end{equation}
Eq. \eqref{eq:exponent_parameter} can be obtained by plugging Eq. \eqref{eq:Meijer_derivative} into Eq. \eqref{eq:CGF_parameter}.

For the tail distribution with $y<\mathbb{E}\{Y_{K_m}\}$, a similar result can be obtained by applying the same process. In this context, Lemma \ref{lem:con_outage_exponent} has been established.

\section{Proof of Lemma \ref{lem:matching_edges}}\label{app:matching_edges}

Let $\mathcal{G}(\mathcal{U}\cup\mathcal{A},\mathcal{E})$ be a bipartite graph with $|\mathcal{U}|=M$ and $|\mathcal{A}|=N$. Clearly, we have $K^\mathrm{sum}\leq NL$ and $K_m\leq N$ for $\forall u_m\in\mathcal{U}$. Suppose $\mathcal{E}$ contains at least one maximum $b$-matching such that the only unsaturated vertex is $u_{j:M}$. According to Lemma \ref{lem:b_matching}, there must be a subset $\mathcal{X}\subseteq\mathcal{U}\cup\mathcal{A}$ satisfying
\begin{equation}\label{eq:matching_cond}
	b(\mathcal{U}\cup\mathcal{A}\setminus\mathcal{X})+|\mathcal{E}(\mathcal{X})|=\sum_{i=1,i\neq j}^M K_{i:M}.
\end{equation}
Denote $\mathcal{X}_\mathcal{U}=\mathcal{X}\cap\mathcal{U}$ and $\mathcal{X}_\mathcal{A}=\mathcal{X}\cap\mathcal{A}$. For any $\mathcal{X}$ which satisfies Eq. \eqref{eq:matching_cond}, there are at most
\begin{equation}\label{eq:matching_edges1}
	(M-|\mathcal{X}_\mathcal{U}|)N+|\mathcal{E}(\mathcal{X}_\mathcal{U})|
\end{equation}
edges if $\mathcal{X}_\mathcal{U}\neq\emptyset$ or
\begin{equation}\label{eq:matching_edges2}
	M(N-|\mathcal{X}_\mathcal{A}|)+|\mathcal{E}(\mathcal{X}_\mathcal{A})|
\end{equation}
edges if $\mathcal{X}_\mathcal{A}\neq\emptyset$. 

Consider Eq. \eqref{eq:matching_edges1} first. As $|\mathcal{E}(\mathcal{X}_\mathcal{U})|\leq |\mathcal{X}_\mathcal{U}|N$,
%\begin{equation}
%	|\mathcal{E}(\mathcal{X}_\mathcal{U})|\leq |\mathcal{X}_\mathcal{U}|N,
%\end{equation}
the maximum value of Eq. \eqref{eq:matching_edges1} is achieved if and only if $|\mathcal{X}_\mathcal{U}|=1$. In this case, the only unsaturated vertex $u_{j:M}$ is in $\{u_{i:M}\}_{i=m}^M$. Then, we have $\mathcal{X}_\mathcal{U}=\{u_{m:M}\}$, $\mathcal{X}_\mathcal{A}=\mathcal{A}$, and $|\mathcal{E}(\mathcal{X}_\mathcal{U})|=K_{m:M}-1$. Therefore, the maximum number of edges in this case is given by
\begin{equation}
	\Phi_1=(M-1)N+K_{m:M}-1.
\end{equation}
Consider then Eq. \eqref{eq:matching_edges2}. According to Eq. \eqref{eq:matching_cond}, the maximum value of $N-|\mathcal{X}_\mathcal{A}|$ must be $\left\lceil\frac{K^\mathrm{sum}}{L}\right\rceil-1$. Otherwise, there will be no unsaturated vertices in $\mathcal{U}$. In this case, the only unsaturated vertex $u_{j:M}$ is in $\mathcal{U}$. Then, we have $\mathcal{X}_\mathcal{U}=\mathcal{U}$ and $|\mathcal{E}(\mathcal{X}_\mathcal{A})|=K^\mathrm{sum}-L\left(\left\lceil\frac{K^\mathrm{sum}}{L}\right\rceil-1\right)-1$.
%\begin{equation}
%	|\mathcal{E}(\mathcal{X}_\mathcal{A})|=K^\mathrm{sum}-L\left(\left\lceil\frac{K^\mathrm{sum}}{L}\right\rceil-1\right)-1.
%\end{equation}
Thus, the maximum number of edges is then given by
\begin{equation}
	\begin{aligned}
		\Phi_2 = & M\left(\left\lceil\frac{K^\mathrm{sum}}{L}\right\rceil-1\right)+K^\mathrm{sum}-L\left(\left\lceil\frac{K^\mathrm{sum}}{L}\right\rceil-1\right)-1 \\
		= & (M-L)\left(\left\lceil\frac{K^\mathrm{sum}}{L}\right\rceil-1\right)+K^\mathrm{sum}-1.
	 \end{aligned}
\end{equation}
In this context, Lemma \ref{lem:matching_edges} has been established.

\section{Proof of Theorem \ref{thm:content_outage}}\label{app:content_outage}

Consider first the high SNR regime. The first order approximation of $p_{m:M}^\mathrm{out}(R_{m:M})$ is determined by the lowest power term of $p_{m:M}$. According to Lemma \ref{lem:con_outage_exponent}, $p_{m:M}^\mathrm{con}(R_{m:M}|\mathcal{D}_{K_{m:M},N-\kappa})$ has the same decreasing order as $p_{m:M}^{N-\kappa}$ when $\gamma$ tends to infinity. Then, the first part of Lemma \ref{lem:matching_edges} indicates that if $\kappa$ Fog-APs are outage for $u_{m:M}$ with $\kappa=N-K_{m:M}+1,\ldots,N$, the content outage probability of $u_{m:M}$ for a fixed $\kappa$ is given by
\begin{equation}
	\begin{aligned}
		& \binom{N}{\kappa}p_{m:M}^\mathrm{con}(R_{m:M}|\mathcal{D}_{K_{m:M},N-\kappa})p_{m:M}^\kappa q_{m:M}^{MN-\kappa} \\
		= &  \binom{N}{\kappa}p_{m:M}^\mathrm{con}(R_{m:M}|\mathcal{D}_{K_{m:M},N-\kappa})p_{m:M}^\kappa+\mathsf{O}\left(p_{m:M}^N\right).
	\end{aligned}
\end{equation}
Therefore, the first order approximation of $p_{m:M}^\mathrm{out}(R_{m:M})$ can be wiritten by
\begin{equation}\label{eq:ProofFirstOrder_1}
%	\begin{aligned}
%		& p_{m:M}^\mathrm{out}(R_{m:M})= \\
%		& \sum_{\kappa=N-K_{m:M}+1}^N \binom{N}{\kappa}p_{m:M}^\mathrm{con}(R_{m:M}|\mathcal{D}_{K_{m:M},N-\kappa})p_{m:M}^\kappa \\
%		& +\mathsf{O}\left(p_{m:M}^N\right)
%	\end{aligned}
	p_{m:M}^\mathrm{out}(R_{m:M})=\sum_{\kappa=N-K_{m:M}+1}^N \binom{N}{\kappa}p_{m:M}^\mathrm{con}(R_{m:M}|\mathcal{D}_{K_{m:M},N-\kappa})p_{m:M}^\kappa+\mathsf{O}\left(p_{m:M}^N\right)
\end{equation}
The second part of Lemma \ref{lem:matching_edges} indicates that at most $\Phi_2$ edges exist if there is one user in $\mathcal{U}$ is in outage. Recalling Eq. \eqref{eq:b_matching_problem}, the user $u_m$ will be allocated at least $\eta K_m$ Fog-APs in this case. Therefore, the content outage probability of user $u_m$ is given by
\begin{equation}\label{eq:ProofFirstOrder_2}
%	\begin{aligned}
%		& p_m^\mathrm{out}(R_m) \\
%		= & \binom{MN}{\Phi_2}p_m^\mathrm{con}(R_m|\mathcal{D}_{K_m,\eta K_m})p_m^{MN-\Phi_2}q _m^{\Phi_2} \\
%		= & \binom{MN}{\Phi_2}p_m^\mathrm{con}(R_m|\mathcal{D}_{K_m,\eta K_m})p_m^{MN-\Phi_2} \\
%		& +\mathsf{O}\left(p_m^{MN-\Phi_2+\eta K_m}\right).
%	\end{aligned}
\begin{aligned}
	& p_m^\mathrm{out}(R_m) = \binom{MN}{\Phi_2}p_m^\mathrm{con}(R_m|\mathcal{D}_{K_m,\eta K_m})p_m^{MN-\Phi_2}q _m^{\Phi_2} \\
	= & \binom{MN}{\Phi_2}p_m^\mathrm{con}(R_m|\mathcal{D}_{K_m,\eta K_m})p_m^{MN-\Phi_2}+\mathsf{O}\left(p_m^{MN-\Phi_2+\eta K_m}\right).
\end{aligned}
\end{equation}
Therefore, if $N<MN-\Phi_2+\eta K_m$, i.e., $N>\frac{\Phi_2-\eta K_m}{M-1}$, the first order approximation is given by Eq. \eqref{eq:ProofFirstOrder_1}. On the other hand, if $N>MN-\Phi_2+\eta K_m$, i.e., $N<\frac{\Phi_2-\eta K_m}{M-1}$, the first order approximation is given by Eq. \eqref{eq:ProofFirstOrder_2}. Finally, if $N=MN-\Phi_2+\eta K_m$, i.e., $N=\frac{\Phi_2-\eta K_m}{M-1}$, the first order approximation is given by the summation of Eq. \eqref{eq:ProofFirstOrder_1} and Eq. \eqref{eq:ProofFirstOrder_2}.

Consider then the low SNR regime such that the sample of $\mathscr{G}\left(\mathcal{K}_{MN};\mathsf{P}\right)$ has a few edges. For a user $u_{m}\in\mathcal{U}$, there are two cases that make $u_{m}$ unsaturated: 1) There are no $K_m$ non-outage Fog-APs in $\mathcal{A}$ for $u_m$; and 2) There are other users competing for the same Fog-AP with $u_m$ and it is not saturated by the maximum $b$-matching. In the first case, the occurrence probability of this event is given by
\begin{equation}
%	\begin{aligned}
%		& \sum_{\kappa=N-K_m+1}^N\binom{N}{\kappa}p_m^\mathrm{con}(R_m|\mathcal{D}_{K_m,N-\kappa})p_m^\kappa q_m^{N-\kappa} \\
%		= & p_m^N+Np_m^\mathrm{con}(R_m|\mathcal{D}_{K_m,1})p_m^{N-1}q_m+\mathsf{O}(q_m),
%	\end{aligned}
	\sum_{\kappa=N-K_m+1}^N\binom{N}{\kappa}p_m^\mathrm{con}(R_m|\mathcal{D}_{K_m,N-\kappa})p_m^\kappa q_m^{N-\kappa} = p_m^N+Np_m^\mathrm{con}(R_m|\mathcal{D}_{K_m,1})p_m^{N-1}q_m+\mathsf{O}(q_m),
\end{equation}
for $\gamma\to0$. In the second case, there will be at least two edges in the bipartite graph $\mathcal{G}(\mathcal{U}\cup\mathcal{S},\mathcal{E})$. One is $e_{mn}=u_ma_n$, and the other one is $e_{m'n}u_{m'}a_n$. Assuming that there are only two edges in the sample of this random bipartite graph, i.e., $\mathcal{E}=\{e_{mn},e_{m'n}\}$. In the maximum $b$-matching, $u_{m}$ or $u_{m'}$ is chosen with equal probability. The outage probability of $u_{m}$ must have the following term
\begin{equation}
	\frac{1}{2}\binom{M}{2}\binom{N}{1}p_m^\mathrm{con}(R_m|\mathcal{D}_{K_m,1})p_m^{MN-2}q_m^2=\mathsf{O}(q_m),
\end{equation}
for $\gamma\to0$. It can be seen that if there are more than two edges in this random bipartite graph, the outage probability of $u_{m}$ must have a factor $q_m^t$ with $t\geq3$. Hence, Eq. \eqref{eq:outage_low} holds.

% Can use something like this to put references on a page
% by themselves when using endfloat and the captionsoff option.
\ifCLASSOPTIONcaptionsoff
  \newpage
\fi

% trigger a \newpage just before the given reference
% number - used to balance the columns on the last page
% adjust value as needed - may need to be readjusted if
% the document is modified later
%\IEEEtriggeratref{8}
% The "triggered" command can be changed if desired:
%\IEEEtriggercmd{\enlargethispage{-5in}}

% references section

% can use a bibliography generated by BibTeX as a .bbl file
% BibTeX documentation can be easily obtained at:
% http://www.ctan.org/tex-archive/biblio/bibtex/contrib/doc/
% The IEEEtran BibTeX style support page is at:
% http://www.michaelshell.org/tex/ieeetran/bibtex/
\bibliographystyle{IEEEtran}
% argument is your BibTeX string definitions and bibliography database(s)
\bibliography{IEEEabrv,library}
\end{document}